\documentclass[fleqn,usenatbib,usedcolumn]{mnras}
\usepackage[british]{babel}             % British English hyphenation
\usepackage{newtxtext}                  % Good fonts
\let\la=\lesssim % for less than similar from newtxmath, not \la from mnras.cls
\usepackage[T1]{fontenc}                % Good font encoding
\usepackage{graphicx}                   % Including figures
\usepackage{color,amsmath,amssymb,mathastext,nicefrac}
\usepackage{lmodern,upgreek}
\usepackage{booktabs}
\usepackage{hyperref}

\title[Lyman Continuum Escape in Tololo 1247-232]{The Lyman Continuum Escape and ISM properties in Tololo 1247-232 -- New Insights from HST and VLA}

\author[J. Puschnig et al.]{
J. Puschnig$^{1}$\thanks{E-mail: johannes.puschnig@astro.su.se},
M. Hayes$^{1}$,
G. \"Ostlin$^{1}$,
T. E. Rivera-Thorsen$^{1}$,
J. Melinder$^{1}$,
\and J. M. Cannon$^{2}$,
V. Menacho$^{1}$,
E. Zackrisson$^{3}$,
N. Bergvall$^{3}$
and E. Leitet$^{3}$
\\
$^{1}$Department of Astronomy, Oskar Klein Centre, Stockholm University, AlbaNova University Centre, 106 91 Stockholm, Sweden\\
$^{2}$Department of Physics \& Astronomy, Macalester College, 1600 Grand Avenue, Saint Paul, MN 55105, USA\\
$^{3}$Department of Physics and Astronomy, Uppsala University, L\"agerhyddsv\"agen 1, 751 20 Uppsala, Sweden
}

\date{Accepted 2017 April 19.}

\pubyear{2017}

\begin{document}
\label{firstpage}
\pagerange{\pageref{firstpage}--\pageref{lastpage}}
\maketitle

\begin{abstract}
Low- and intermediate mass galaxies are widely discussed as cause of reionization at redshift $z\sim10-6$.
However, observational proof of galaxies that are leaking ionizing radiation (Lyman continuum; LyC) is
a currently ongoing challenge and the list of LyC emitting candidates is still short. Tololo 1247-232
is among those very few galaxies with recently reported leakage. We performed intermediate resolution
ultraviolet (UV) spectroscopy with the Cosmic Origins Spectrograph (COS) onboard the Hubble Space
Telescope and confirm ionizing radiation emerging from Tololo 1247-232.
Adopting an improved data reduction procedure, we find that LyC escapes from the central stellar
clusters, with an escape fraction of 1.5$\pm$0.5\% only, i.e. the lowest value reported for the galaxy so far.
We further make use of FUV absorption lines of Si II and Si IV as a probe of the
neutral and ionized interstellar medium. We find that most of the ISM gas is ionized, likely
facilitating LyC escape from density bounded regions. Neutral gas covering as a function of
line-of-sight velocity is derived using the apparent optical depth method. The ISM is found to be
sufficiently clumpy, supporting the direct escape of LyC photons. We further report on broadband UV
and optical continuum imaging as well as narrowband imaging of Ly$\upalpha$, H$\upalpha$ and H$\upbeta$.
Using stellar population synthesis, a Ly$\upalpha$ escape fraction of 8\% was derived. We also performed VLA 21cm
imaging. The hydrogen hyperfine transition was not detected, but a deep upper limit atomic gas mass
of $\lesssim10^9 M_{\odot}$ could be derived. The upper limit gas fraction defined as
$\nicefrac{M_{H I}}{M_*}$ is only 20\%. Evidence is found that the H I gas halo
is relatively small compared to other Lyman Alpha emitters.
\end{abstract}


%%%%%%%%%%%%%%%%%%%%%%%%%%%%%%%%%%%%%%%%%%%%%%%%%%

%%%%%%%%%%%%%%%%% BODY OF PAPER %%%%%%%%%%%%%%%%%%

%%%%%%%%%%%%%%%%%%%%%%%%%%%%%%%%%%%%%%%%%%%%%%%%%%%%%%%%%%%%%%%%%%%%%%%
%%%%%%%%%%%%%%%%%%%%%%%%%%%%%%%%%%%%%%%%%%%%%%%%%%%%%%%%%%%%%%%%%%%%%%%
%%%%%%%%%%%%%%%%%%%%%%%%%%%%%%%%%%%%%%%%%%%%%%%%%%%%%%%%%%%%%%%%%%%%%%%
\section{Introduction}
%Ho = 70, OmegaM = 0.300, Omegavac = 0.700, z = 0.048
%The luminosity distance DL is 213.1 Mpc or 0.695 Gly. 
Our current understanding of the cosmological evolution after the Big Bang
is well described within the $\Lambda$CDM model which implies two
ages of major gas phase changes. The recombination, caused due to
the rapid cooling of the Universe after $\sim400$ thousand years at $z\sim1090$,
and the reionization, that gradually ionized the Universe
until $z\sim6$ \citep{Fan2001,Fan2006} corresponding to an age of $\sim1$\,Gyr.
As cause of reionization several mechanisms are discussed, such as 
accreting black holes or Pop III stars \citep{Madau2004,Trenti2009,Volonteri2009,Mirabel2011,Finlator2016},
but low- and intermediate mass galaxies seem to be the top candidate,
given that some fraction of their ionizing radiation was able to escape into the intergalactic
medium.
However, most of the galaxies observed yet did not show any signs of such Lyman continuum (LyC) emission,
but upper limits could be derived by various authors: e.g. \citet{Leitherer1995,Deharveng2001}
for galaxies in the local Universe and \citet{Siana2007,Cowie2009,Rutkowski2015} for galaxies at $z\sim1$.
Nevertheless, LyC escape of galaxies at $z\sim3$ was reported by \citet{Iwata2009,Nestor2011,Mostardi2015,Micheva2015},
but could be suffering from foreground contamination \citep{Siana2007,Vanzella2010}. So far, the most convincing cases
of LyC leakage in individual galaxies at high redshift were recently published by \citet{Vanzella2016} and
\citet{deBarros2016}.

Local galaxies that can be observed at high resolution, exhibiting 
LyC leakage thus may shed light onto the mechanisms that
allow LyC photons to escape.
However, despite many observational efforts have been made in the past to detect escaping
LyC radiation in galaxies at all redshifts \citep{Bergvall2013},
%star forming galaxies beyond the local universe fall into two categories. At z ≈ 1, space based
%ultraviolet (UV) imaging or spectroscopy around the Lyman break has been used to infer
%escaping ionizing radiation (e.g., Bridge et al. 2010; Rutkowski et al. 2016). No
%conclusive detection has been made to date (Siana et al. 2015). At higher redshift (z ≈ 2 –
%3), both space-based and ground-based observations are feasible. Evidence of Lyman
%continuum emission must be taken with care due to possible contamination by low-
%redshift interlopers (Mostardi et al. 2015). Nevertheless, Lyman continuum radiation has
%been detected in about 10% of the galaxies surveyed (Siana et al. 2015), with escape
%fractions ranging between 5% and 30% for Lyman-break galaxies and Lyman-α emitters,
%respectively (Nestor et al. 2013). Vanzella et al. (2016) reported an exceptional case of a
%star-forming galaxy at z = 3.2 whose escape fraction was found to exceed 50%.
the list of candidates is still short,
in particular at low redshifts where LyC observations are possible only through
space-based observatories such as FUSE, the Far-Ultraviolet Spectroscopic Explorer \citep{Moos2000}
or the Hubble Space Telescope (HST).
To date, only few local galaxies have confirmed LyC detections. First discoveries were made in
Haro11 \citep{Leitet2011}, Mrk 54 \citep{Leitherer2016},
Tololo 1247-232 \citep{Leitet2013,Leitherer2016} and J0921+4509 \citep{Borthakur2014}.
The published LyC escape fractions are 3.3$\pm$0.7\% (Haro11), 2.5$\pm$0.72\% (Mrk 54), 4.5$\pm$1.2\% (Tololo 1247-232)
and 21$\pm$5\% (J0921+4509).
Recent publications by \citet{Izotov2016a,Izotov2016b} further add five other sources to the list,\ 
with the highest measured escape fraction found in J0925+1403 (7.8$\pm$1.1\%). 

Yet, in order to understand if galaxies could have reionized the Universe, we need
to find links between readily accessible observables and the
escape of ionizing photons. That way, the number of known LyC leakers could be drastically increased,
allowing to disentangle the complex physical processes driving LyC escape.
Work done by \citet{Zastrow2011,Zastrow2013} suggest that so called \textit{indirect tracers} such as ionization-parameters
can provide such a link. That followed, \citet{Jaskot2013} suggest that a high [O\,III]$\uplambda$5007/[O\,II]$\uplambda$3727 ratio
as typically observed in Green Pea galaxies could give rise to leakage of ionizing radiation,
because optically thin nebulae should underproduce [O II] \citep{Pellegrini2012}. Hence,
high [O\,III]$\uplambda$5007/[O\,II]$\uplambda$3727 ratios may also indicate a low LyC optical depth.
Another indirect approach was described by \citet{Heckman2011}, who argue that
\textit{non-saturated optically thick} FUV metal
absorption lines of C\,II or Si\,II can be used as a proxy for Lyman continuum, since residual line flux
along the absorption spectrum may indicate the presence of channels with low column densities within
a medium filled with optically thick clouds, i.e. in agreement with the picket-fence model \citep{Heckman2001}.
Since typical H\,I column densities in galaxies are easily high enough to prevent LyC escape,
in the picket-fence model, starbursts could create a sufficiently porous ISM allowing
the ionizing radiation from the massive stars to escape through low density areas or voids.
Another spectral line discussed in various previous studies that may give rise to Lyman continuum emission,
is escaping Ly$\upalpha$ radiation. This is particular interesting, because
coupling between Ly$\upalpha$ and LyC \citep{Verhamme2015} would further mean a big step forward towards understanding
the importance of Ly$\upalpha$ emitters (LAEs) during the epoch of reionization,
given the large and increasing number of LAEs observed in the early Universe,
exactly as needed for reionization \citep{Stark2010,Curtis-Lake2012}.
In that context the works by e.g. \citet{Nakajima2014} and \citet{Henry2015} seem promising,
since they find that the high Ly$\upalpha$ escape fractions in their samples are related to
a low H\,I column density rather than outflow velocity or H\,I covering fraction.

Luckily, various models and simulations show that the ionizing photon escape
increases with decreasing dark matter halo mass \citep{Yajima2011,Ferrara2013,Paardekooper2013,Wise2014},
making LAEs even more promising to be good candidates for LyC leakage.
Additionally, cosmological hydrodynamic simulations performed by
e.g. \citet{Yajima2014} find a relation between the emergent Ly$\upalpha$ radiation and 
the ionizing photon emissivity of LAEs.
This is further supported by \citet{Dijkstra2016} who performed Monte-Carlo radiative transfer simulations
through models of dusty, clumpy interstellar media. The authors find that the ionizing escape fraction is
strongly affected by the cloud covering factor, which implies that LyC leakage is closely connected to the
observed Ly$\upalpha$ spectral line shape, because
the escape of ionizing photons requires that sightlines with low column densities exist,
i.e. $N_{H I}\lesssim10^{17}\,cm^{-2}$. These same channels of low column density then also provide
escape routes for Ly$\upalpha$ photons \citep{Behrens2014,Verhamme2015}; leading to
Ly$\upalpha$ halos that are comparably small.

\begin{figure*}
\centering
   \includegraphics[width=18cm]{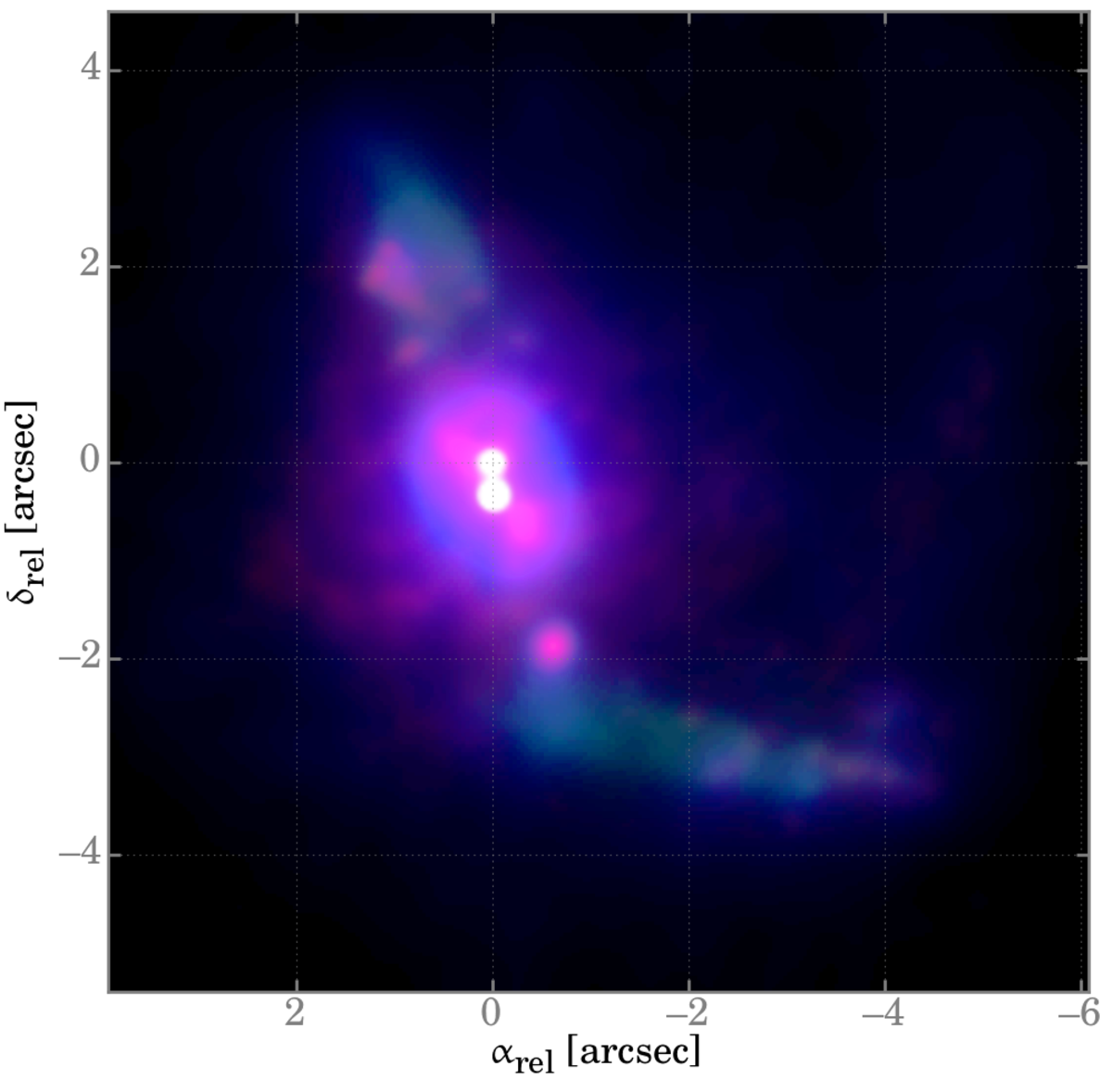}
     \caption{RGB false-color composite showing H$\upalpha$ in $red$, Ly$\upalpha$ in $blue$ and the UV
     continuum in $green$. Each plane was scaled and smoothed individually to emphasize on morphological
     differences between the individual components. The coordinates are relative to the center of
     the COS aperture. North is up, east is left. }
     \label{fig:rgb}
\end{figure*}

In this paper, using our newly obtained HST FUV spectra of Tololo 1247-232
(see basic parameters in Table \ref{tab:basicprop}), along with direct observations of the Lyman continuum, we
are able to test the previously described method of \citet{Heckman2011} through FUV Si~II
absorption lines. Our observation of the Ly$\upalpha$ line
allows us to discuss coupling between LyC and Ly$\upalpha$ as proposed by
\citet{Verhamme2015} and using VLA 21\,cm data,
we are further able to make constraints on the atomic gas reservoir of Tololo 1247-232.
Additionally, we have obtained HST narrowband imaging.
A false-color composite of the galaxy encoding the morphology of Ly$\upalpha$, H$\upalpha$ and UV continuum is
shown in Figure \ref{fig:rgb}.
Given the amount of high quality data now available, we eventually discuss the driving
mechanisms and physical quantities behind the LyC of Tololo 1247-232 (hereafter Tol1247).
Throughout the paper, we adopt a cosmology with H$_0 = 70$, $\Omega_M = 0.3$ and $\Omega_{vac} = 0.7$.
The measured redshift of Tol1247 is $z=0.0488\pm0.0002$\footnote{based on optical emission lines using NOT/ALFOSC slit spectra},
which results in a luminosity distance of 216.8\,Mpc.
\begin{table}
\centering
\begin{tabular}{l c c}
\toprule
parameter & value & reference \\
\midrule
 redshift              & $z = 0.0488\pm0.0002$                             & see caption \\
 luminosity dist.      & $216.8\,Mpc$                                      & from $z$ (W06) \\
 stellar mass          & $M_* = 5 \times 10^9\ M_{\odot}$                  & L13 \\
 atomic gas mass       & $M_{H I} \lesssim 10^9\ M_{\odot}$                & this work \ref{sec:atomicgas} \\
 gas fraction          & $\nicefrac{M_{H I}}{M_*} \lesssim 0.2$            & this work \ref{sec:halosize} \\
 star formation rate   & SFR$_{H\upalpha,0,SB} = 36.2\,M_{\odot}\,yr^{-1}$ & this work \ref{sec:SFR}  \\
 metallicity           & $12+log\big(\nicefrac{O}{H}\big) = 8.1\ or\ \approx\nicefrac{1}{4} Z_{\odot}$ & T93 \\
\bottomrule
\end{tabular}
\caption{Basic parameters of Tololo 1247-232. The redshift was measured from optical emission lines
by Emily Freeland (Stockholm University) using ALFOSC slit spectra obtained at the Nordic Optical Telescope.
Reference codes are W06 for Ned Wright's cosmology calculator \citep{Wright2006},
L13 is \citet{Leitet2013} and T13 is \citet{Terlevich1993}.}
\label{tab:basicprop}
\end{table}

%%%%%%%%%%%%%%%%%%%%%%%%%%%%%%%%%%%%%%%%%%%%%%%%%%%%%%%%%%%%%%%%%%%%%%%
%%%%%%%%%%%%%%%%%%%%%%%%%%%%%%%%%%%%%%%%%%%%%%%%%%%%%%%%%%%%%%%%%%%%%%%
%%%%%%%%%%%%%%%%%%%%%%%%%%%%%%%%%%%%%%%%%%%%%%%%%%%%%%%%%%%%%%%%%%%%%%%
\section{Observations and Data Reduction}
% PI: G\"oran \"Ostlin
% PropID: 13027
% TIME-TAG, PSA(2.5arcsec), G130M
%
% DATE-OBS: 2013-05-29
% CENWAVE=1055\AA
% EXPTIME=1020s,1020s,1307s,1307s
% FP-POS 1-4: LBZ911O2Q, LBZ911O4Q, LBZ911O6Q, LBZ911O8Q
%
% DATE-OBS: 2013-07-18
% CENWAVE=1055\AA
% EXPTIME=1020s,1020s,1307s,1307s
% FP-POS 1-4: LBZ912T0Q,LBZ912T2Q,LBZ912T4Q,LBZ912T6Q
%
% DATE-OBS: 2013-07-18
% CENWAVE=1327\AA
% EXPTIME=588s,588s,588s,588s
% FP-POS 1-4: LBZ912T8Q,LBZ912TAQ,LBZ912TCQ,LBZ912TEQ
%
%%%%%%%%%%%%%%%%%%%%%%%%%%%%%%%%%%%%%%%%%%%%%%%%%%%%%%%%%%%%%%%%%%%%%%%%%5
\subsection{HST/COS FUV Spectroscopy} \label{sec:calcosmod}
FUV spectroscopy was performed using the Cosmic Origins Spectrograph (COS) at the Hubble Space Telescope (HST)
during May and July 2013. The observations were carried out in TIME-TAG mode using the G130M grating
in two different wavelength regions, the first with a central wavelength of 1055\,\AA\ sampling the full wavelength range to target LyC and the second
with a central wavelength of 1327\,\AA\ targeting the Ly$\upalpha$ as well as a multitude of ISM absorption lines.
In order to minimize effects of small scale fixed pattern noise in the detector, individual exposures
were taken using the four available grating offset positions (FP-POS) that move the spectrum slightly
in the dispersion direction and allow the spectrum to fall on different areas of the detector.
The detector consists of two 16384x1024 pixel segments, referred to as FUV segments A and B, or FUVA and FUVB,
leaving a small gap inbetween.
The total exposure times were 9308s ($\approx$2.6h) and 2352s ($\approx$0.7h) for the LyC and Ly$\upalpha$ part respectively.
We have used the circular primary science aperture (PSA) of 2.5" diameter centered on the galaxy's brightest cluster (as seen at 1500\,\AA).
Since COS is an aperture spectrograph, the spectral resolution $R$ depends on the source's extent/distribution convolved with the instrumental
line spread function (LSF). The COS manual gives $R\approx20000$ when using grating G130M on a point source.
Due to the compactness of the UV continuum emission, the metal absorption line observations such as e.g.
Si\,II\ (see Figure \ref{fig:spectrum1327both}) will thus have a comparably high spectral resolution. However, for a source totally filling the aperture,
the spectral resolution decreases to $R\approx1500$. This value is most likely true
for the scattered Ly$\upalpha$ emission in Tol1247.

The data reduction was performed using a modified version of \texttt{CALCOS 2.21}.
With the new pipeline, the noise level could be significantly decreased, especially
in the FUVB below $\approx$1100\,\AA, where the COS throughput changes by a factor of
$\approx$100; see section 5.1.2 of the COS instrument handbook \citep{COSIHB}.
In the following the improvements of the new pipeline are described. See also Justin Ely's
website\footnote{\url{https://justincely.github.io/AAS224/}}. Some of his code examples
were used as a template.

\subsubsection{Adjusting the Pulse Height Amplitude Limits}
With each registered photon event at the detector, a scaled pulse height amplitude (PHA) value between 0 and 31
is saved. Interestingly, noise events typically have very large or very small PHAs.
Hence, filtering events using PHA limits can be used to reduce the noise in the final spectrum.
Although \texttt{CALCOS 2.21} already adopts such limits (4 and 30), using tailored values for individual
science exposures can still improve the noise reduction.
For our Tol1247 data, we found lower and upper PHA thresholds of 2 and 18 respectively, which slightly
enhances the final SNR, especially in the FUVB part of the spectrum.

\subsubsection{Sun altitude}
The following data filtering process was previously described by \citet{James2014}. Here,
we only briefly review the basic concept and refer to \citet{James2014} for more details.
Spectra obtained with the HST are subject to contamination from geocoronal emission lines.
The two strongest airglow features are the Ly$\upalpha$ line at 1215.67\,\AA\ and the
O I feature between 1302\,\AA\ and 1306\,\AA, but a multitude of other weak airglow lines is present in the
UV that add up to the background level; see \citet{Feldman2001} for a compilation of airglow lines.
These lines strongly vary with the solar activity and the altitude of the sun above the horizon --
in case of the HST the geometric horizon. Hence it can be advantageous to use only those photon events
that were registered when the sun altitude was below a certain limit (this information is stored in the so called
"TIMELINE" extension of the binary fits table with the suffix "corrtag").
Since filtering data in that way at the same time also reduces the exposure time, this strategy
might not always improve the result. However, for our Tol1247 exposures we found that using only data
with an altitude threshold of 45$^\circ$ significantly lowers the noise level.
%the "GTI" extension in the "corrtag" file was used to define the good time intervals when the sun was below the given threshold
%Finally, the SNR in the extracted spectrum could be significantly enhanced by this step.

\subsubsection{Supderdark}
Background correction as performed by the unmodified version of \texttt{CALCOS 2.21},
uses a module that estimates the background contribution to the extracted spectrum and
subtracts it from the 1D science spectrum. These estimates are based on
computations done using two predefined regions above and below the spectral extraction
region. However, the actual background level at the spectral position can
differ from the background at the locations used for computing the background.
This leads to either over- or under-subtraction of the background/dark current in the final spectrum.
Hence, an optimized and accurate background correction is needed for faint sources.
This task was performed using a modified pipeline version, \texttt{CALCOS 2.21d}
(C. Leitherer and S. Hernandez priv. comm; \citet{Leitherer2016}).
The new version subtracts a superdark image from the science exposure
right before the 1D extraction.
The superdark has to be provided by the user as a new reference file.
We created this file by combining dark frames from the HST archive (FUV dark monitoring program).
Only those darks were chosen that were imaged two weeks before and after the science exposures.
In that way, the influence of observed systematic variations of the dark current was minimized.
In a final step, the superdark was smoothed using a boxcar filter of size (10,100). This is crucial
to find averaged dark current levels for all pixels.

\subsubsection{Scaling the Superdark and Spectral Extraction}
We have further modified \texttt{CALCOS 2.21d} and implemented the following features:

\begin{itemize}
\item{The superdark is created on-the-fly from an archive of individual dark frames.}
\item{The mean count rates from both nominal background regions are used to scale the superdark to the science exposure. This makes the dark levels of the superdark and the science frame comparable to each other.}
\item{The scaled superdark is then subtracted from each science frame. Before subtraction, the science frame can be smoothed.}
\item{The central line of the spectral extraction region can be placed at any region of the detector.}
\item{Relative to the central line, an extraction size in pixel can be chosen, i.e. different extraction sizes can be used.}
\item{A slightly modified routine to calculate the final extracted net count rate (from which the flux is derived) was implemented. This was needed since \texttt{CALCOS 2.21d} could not correctly handle negative count rates (see appendix \ref{sec:appendix} for more details).}
\end{itemize}

%For our Tol1247 data we have used boxcar smoothing on the LyC part of the spectrum over 2 pixel along the dispersion axis,
%before the 1D spectrum is extracted (which slightly degrades the spectral resolution). Additionally, the extraction procedure was modified to
%automatically scale the provided superdark file to the background level of the given science frame.
%This was done using the count rates within the predefined standard background regions as used by
%the unmodified \texttt{CALCOS} version. After scaling the superdark and smoothing the science frame
%and subtracting the former one from the latter, the 1D extraction was finally performed.
The individual extracted spectra (including different FP-POS) were then combined to a final spectrum using the \texttt{IRAF/PyRAF} task
\texttt{splice} and a scalar weighting based on the effective exposure time (after filtering data according to the sun altitude threshold).
The final spectra were brought on a equi-distant wavelength vector, using a binning of 0.01\AA\ for all but the LyC part, where a binning of 0.5\AA\ was used.

\subsubsection{Error Estimate}
The errors reported by \texttt{CALCOS} are based on Gehrels' variance function \citep{Gehrels1986}, which
offers only an upper limit for the uncertainty regardless of the number of source counts \citep{Fox2014,COSDHB}.
In order to derive a more realistic error estimate, we experimented with spectral extractions of different locations of the detector, e.g. along
the nominal background regions or the different lifetime positions. Without superdark correction, we could identify systematic differences
(in the noise level) between the two nominal background regions, which disappear once the superdark subtraction is performed.
Finally, we choose to use the noise measured in a set of individual dark frames,
that were scaled to our science frames and further treated in exactly the same manner as our set of science frames. The noise was then
evaluated exactly at the detector location of the science spectrum, using the same aperture size.
\begin{figure*}
\centering
   \includegraphics[width=18cm]{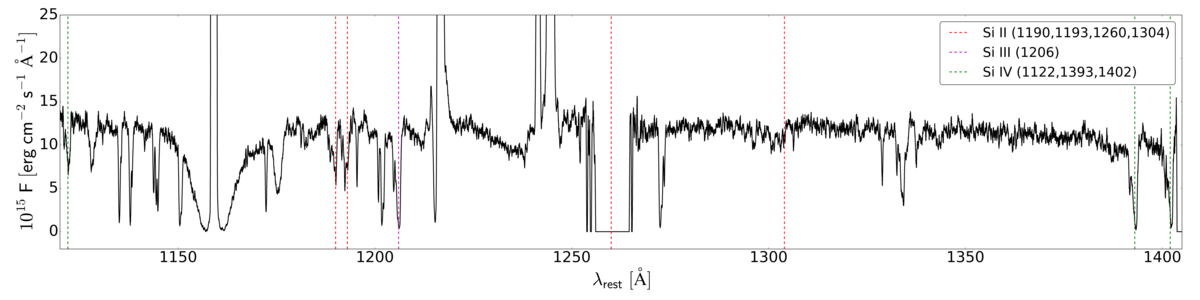}
     \caption{Cutout of the reduced COS spectrum showing the
     relevant metal absorption lines of Si\,II and Si\,IV. Note that the Si\,II absorption line at 1260\AA\ falls into the
     gap between the detector segments A and B.}
     \label{fig:spectrum1327both}
\end{figure*}

%%%%%%%%%%%%%%%%%%%%%%%%%%%%%%%%%%%%%%%%%%%%%%%%%%%%%%%%%%%%%
%%%%%%%%%%%%%%%%%%%%%%%%%%%%%%%%%%%%%%%%%%%%%%%%%%%%%%%%%%%%%
\subsection{HST Imaging}
Tol1247 was imaged with the HST in the optical
using the Wide Field Ultraviolet-Visible Channel (UVIS) of its Wide Field Camera~3 (WFC3).
For UV imaging the Advanced Camera for Surveys' Solar Blind Channel (ACS/SBC) was used.
Seven filters were
utilized in total, allowing to apply LaXs -- the Lyman alpha eXtraction software \citep{Hayes2009} -- to produce
continuum-subtracted Ly$\upalpha$, H$\upalpha$ and H$\upbeta$ images, corrected for underlying stellar
absorption and contamination from [N\,II]$\uplambda6548,6584$.
The latter one is based on the spectroscopic line ratio $\nicefrac{[N\,II]}{H\upalpha}=0.0605$
published by \citet{Terlevich1993}.
The imaging strategy and data reduction methodology for Tol1247 is very similar to that of
the Lyman Alpha Reference Sample (LARS; \citet{Hayes2014,Ostlin2014,Duval2015}) and the basic data
reduction for this data set is done in the same way as for LARS.
Flatfield-corrected frames were obtained from the Mikulski
Archive for Space Telescope. The Charge Transfer Inefficiency (CTE) correction for the
ACS data was performed by the pipeline, whereas CTE losses in WFC3/UVIS \citep{Anderson2010}
were treated manually using the tools supplied by STScI\footnote{\url{http://www.stsci.edu/hst/wfc3/tools/cte_tools}}.
We then stacked the individual data frames and drizzled them to a pixel scale of 0.04\,arcsec/px using
DrizzlePac version 1.1.16 \citep{Gonzaga2012}. Further pre-processing of the data includes additional
masking of cosmic rays in the drizzled frames and matching the point spread functions (PSFs) for the
different filters.
In order to match the PSF, we first construct PSF models for all of the filters used in
the study. For the optical filters we use TinyTim models \citep{Krist2011}, resampled to a
pixel scale of 0.04\,arcsec/px. However, for the FUV filters TinyTim is not accurate enough, in particular in the wings.
Therefore, the PSF models for the FUV filters are instead built from stacks of stars obtained in calibration
observations; see \citet{Hayes2016} for details. All of the PSF models are then normalized by
peak flux and stacked by maximum pixel value. We then proceed to calculate convolution kernels that
match the PSFs for all of the filters to the maximum width model. Each kernel is built up from a
set of delta functions and we find the optimum matching kernel by least squares optimization;
see also \citet{Becker2012} and Melinder et al. (in prep). The drizzled and registered images are convolved with the kernel
found for each filter, which result in a set of images matched to a common PSF.

\begin{figure*}
\centering
   \includegraphics[width=18cm]{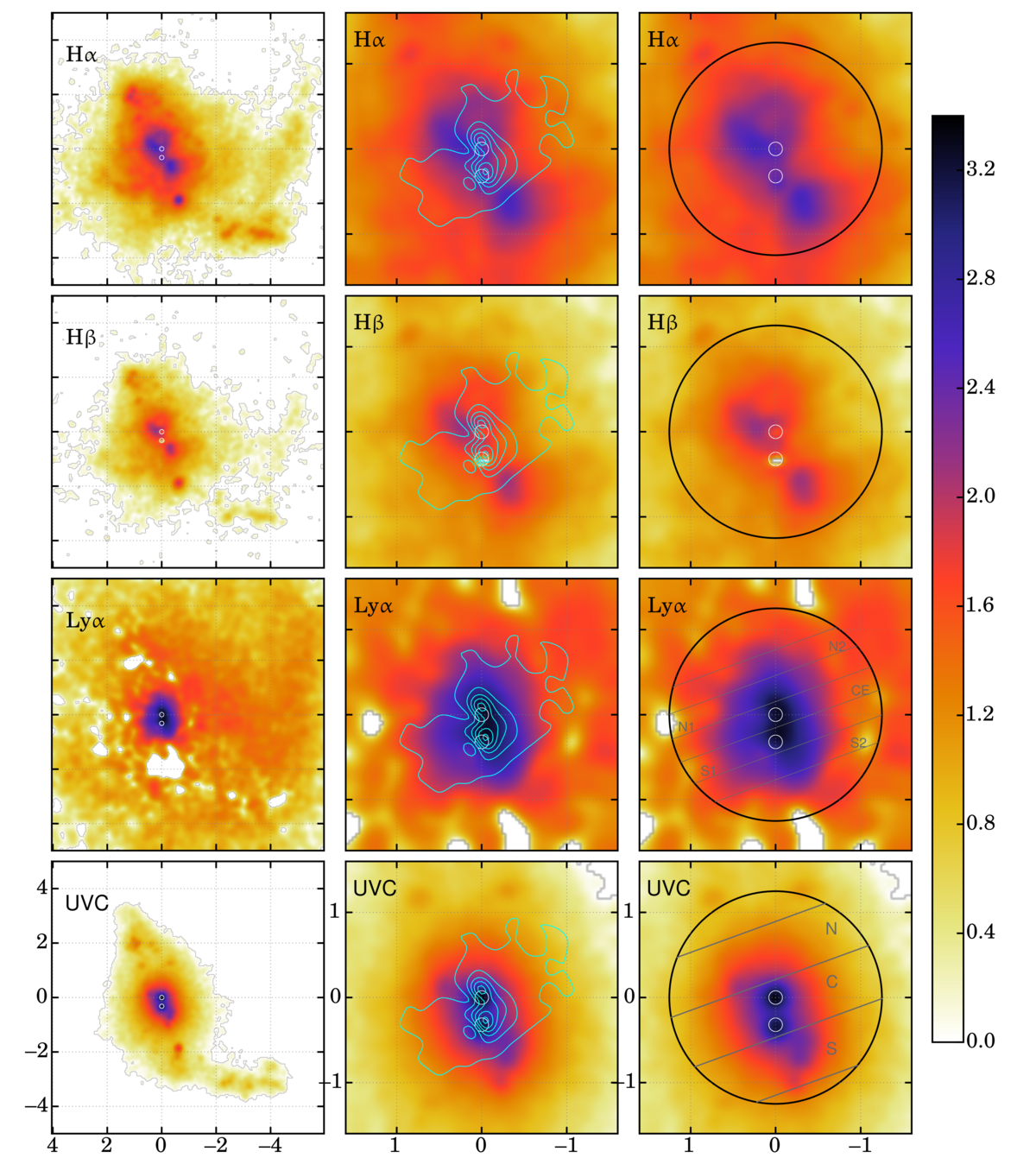}
     \caption{Results from HST imaging. North is up, east is left, with relative coordinates in arcsec.
     Images shown in the left column have a field of view of 10"x10", the remaining ones are zoomed-in versions
     that cover the central $\sim$3"x3".
     The logarithmic color scale is the same for all images and ranges from 0 to $10^{3.4}$ in units of
     $10^{-18} erg\ s^{-1}\ cm^{-2}$ for H$\upalpha$, H$\upbeta$, Ly$\upalpha$ and units of $10^{-19} erg\ s^{-1}\ cm^{-2}\ \AA^{-1}$ for the
     UV continuum (UVC). The black circle has a diameter of 2.5~arcsec, the size of the COS aperture, corresponding to 2.6~kpc.
     The cyan contours show the (smoothed) line ratio of $\nicefrac{Ly\upalpha_{obs}}{H\upalpha_0}$ with
     $H\upalpha_0$, the extinction corrected line flux. Thus, they trace regions of increasing Ly$\upalpha$
     escape fractions. Contour levels are: 0.5, 1.5, 2.5, 3.5, 4.5 and 5.5 -- corresponding to escape fractions
     between roughly 5 and 55\%. The positions of the two clusters as seen in the UV are overplotted as white circles.
     The Ly$\upalpha$ and UVC images in the right column further show individual spectral extraction regions
     (the position angle of the COS PSA was -20$^{\circ}$).}
     \label{fig:hstmaps}
\end{figure*}

%%%%%%%%%%%%%%%%%%%%%%%%%%%%%%%%%%%%%%%%%%%%%%%%%%%%%%%%%%%%%
%%%%%%%%%%%%%%%%%%%%%%%%%%%%%%%%%%%%%%%%%%%%%%%%%%%%%%%%%%%%%
\subsection{Lyman alpha eXtraction software}
A galaxy's spectral energy distribution (SED) can be understood as the
sum of contributions from different components: a starburst (current episode of star formation),
an underlying old stellar population (field stars) and nebular emission (continuum and lines).
In this study, we make use of LaXs \citep{Hayes2009}, the Lyman alpha eXtraction software,
which builds up upon the Starburst99 \citep{Leitherer1999} spectral evolutionary models.
As input for the models we assume a Kroupa IMF \citep{Kroupa2001} and a SMC extinction law.
The models are then constrained through SED fitting using data from our HST imaging in on-line and off-line filters.
As a result, stellar ages, the star-formation history, extinction E(B-V) and the ionizing flux could be derived
for each pixel.

Continuum subtraction plays a critical role and is performed using a dimensionless quantity,
the continuum throughput normalization (CTN),
to scale the raw count rate of the observed continuum flux (off-line) to that expected in the on-line filter.
Thus, CTN is the only quantity needed when performing continuum subtraction. However, it cannot be
estimated without some knowledge of the continuum itself. As shown in \citet{Hayes2005}, degeneracies such
as the one between age and extinction, can be resolved using (at least) four HST broadband (off-line) filters,
chosen to sample the UV slope $\upbeta$, which is sensitive to both age and dust, as well as the Balmer break
at 4000\,\AA, that is primarily sensitive to the age of the stellar population.

Based on the constrained starburst population, LaXs also creates a map of the intrinsic flux density at
900\,\AA\ from which the expected average flux density within apertures of varying sizes centered on the position
of the COS aperture could be derived. The LyC escape fraction (see section \ref{sec:LyCescape}) is then
defined as the ratio between the predicted and observed flux density at 900\,\AA. 
%%%%%%%%%%%%%%%%%%%%%%%%%%%%%%%%%%%%%%%%%%%%%%%%%%%%%%%%%%%%%%%%%%%%%%%%%5
%%%%%%%%%%%%%%%%%%%%%%%%%%%%%%%%%%%%%%%%%%%%%%%%%%%%%%%%%%%%%%%%%%%%%%%%%
\subsection{VLA Data}
\begin{figure}
\centering
   \includegraphics[width=1\columnwidth]{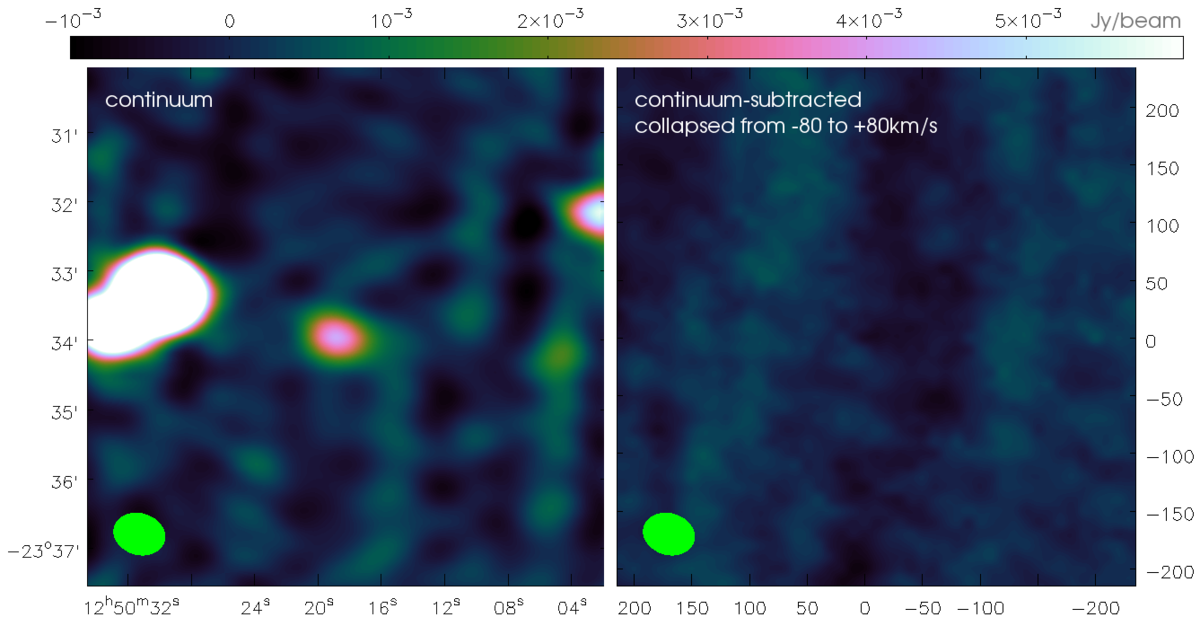}
     \caption{Cutout of the VLA 1.4\,GHz image. The continuum subtraction was performed using \texttt{CASA}'s task \texttt{imcontsub}.
     Tololo 1247-232 is seen in the continuum (\textit{left panel}) as unresolved source
     in the center of the image. The two nearby sources are NVSS J125030-233323 and J125001-233210 to its left and right
     respectively. The synthesized beam size of 45"x36" is shown in the lower left corner as green ellipse.
     The hydrogen 21\,cm line was not detected. After fitting a 0-order polynomial to the continuum at each pixel
     and collapsing the spectrum from -80 to +80\,km/s relative the nominal position of the line, only noise is
     left (\textit{right panel}).}
     \label{fig:vlaimage}
\end{figure}

% Proposal ID VLA/14B-194
% PI Matthew Hayes
% Where is the H I in the Strongest Low Redshift Lyman Continuum Emitting Galaxy?
% 5.8 hours
% Continuum Flux: 
% Gain Calibrator: J1248-1959
% Bandpass Calibrator: 1331+305=3C286
% 27 antennas in DnC configuration
% 2048 channels
% Total BW: 32MHz
% ChanWidth: 15.625kHz
% Chan0: 1336.226MHz
% Tol1247: Vsys: 14390km/s
% Restfreq: 1420.405752
% Synthesized beam size using natural weighting: ['45arcsec','36arcsec','74.0deg']
Tol1247 was observed at 1.4\,GHz with the Karl G. Jansky Very Large Array (VLA) under
program 14B-194 in September 2014 aiming to quantify
the atomic gas mass of the galaxy. The VLA was operated with all 27 antennas
in a hybrid DnC configuration,
in which the north arm antennas are deployed in the next larger configuration
than the SE and SW arm antennas. The hybrid configuration was chosen, because
of the low source declination of -23$^\circ$, for which the extended northern
arm results in a more circular shaped synthesized beam. The total observing time
was roughly 6 hours including overhead. For phase calibration, J1248-1959 was observed
regularly and the flux density scale was fixed by observing one standard
NRAO\footnote{The National Radio Astronomy Observatory is a facility of the National
Science Foundation operated under cooperative agreement by Associated Universities, Inc.}
flux calibrator, 1331+305 (3C286), for which the standard NRAO flux density of $\approx15.5$\,Jy was employed.
The measurement sets were checked for bad baselines which were subsequently flagged,
before self-calibration was performed with \texttt{CASA 4.2.2} using (E)VLA pipeline scripts.
For imaging, we experimented with different levels of cleaning using \texttt{CASA}'s interactive mode.
The cleaning threshold was found after the flux densities of unresolved sources did not further change.
Uniform, natural and Briggs baseline weighting was applied to check sensitivity levels.
Additional uv-tapering circularized the synthesized beam, but did not improve the sensitivity. Finally, the best result was found for natural weights
without uv-tapering (see Figure~\ref{fig:vlaimage}). The synthesized beam size is 45"x36" with a position angle of 74 degree.

\begin{figure}
\centering
   \includegraphics[width=1\columnwidth]{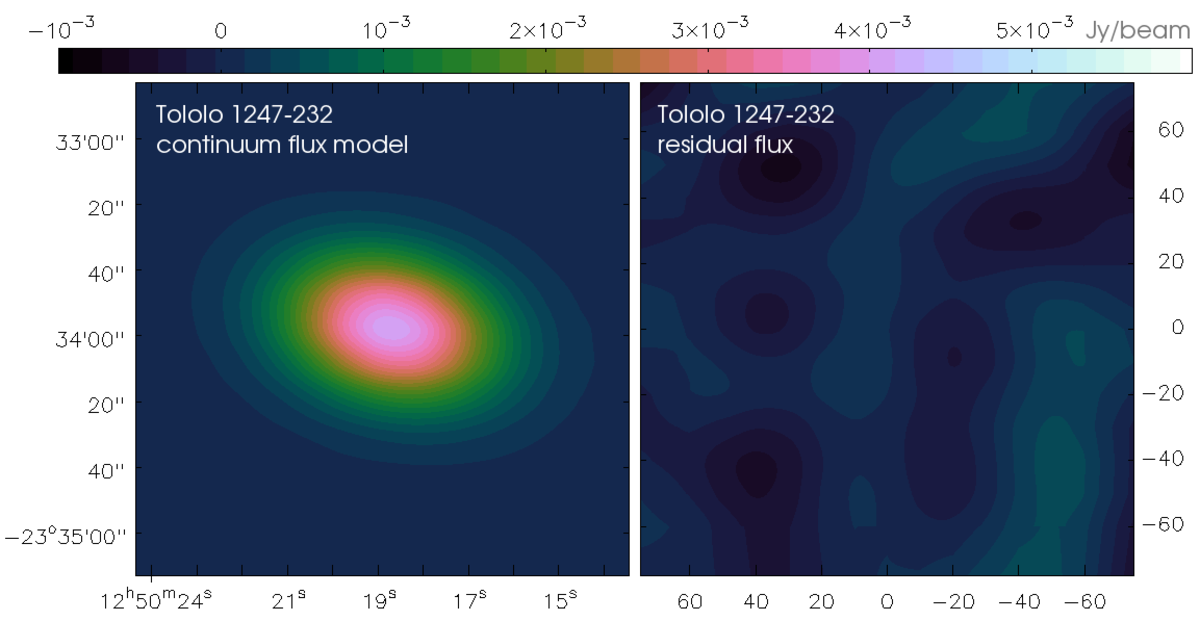}
     \caption{2D Gaussian continuum flux density model of Tololo 1247-232 (\textit{left}) and residual (\textit{right}) after
     subtracting the model from the collapsed continuum image.}
     \label{fig:vlafluxmodel}
\end{figure}

To measure the flux density at 1.4\,GHz, we fit a 2D Gaussian to the collapsed
continuum image as seen in left panel of Figure~\ref{fig:vlaimage} using \texttt{CASA}'s
task \texttt{imfit}. The flux density is then derived from the model (Gauss)
shown in Figure~\ref{fig:vlafluxmodel}. We perform the same task
on two other nearby sources, NVSS J125030-233323 and NVSS J125001-233210 to check
our results against previously published values from \citet{Condon1998} and find
good agreement (see Table~\ref{tab:21cmflux}).

\begin{table}
\centering
\begin{tabular}{l c c}
\toprule
 & S$_{1.4GHz}$ [mJy]  & S$_{1.4GHz}$ [mJy]   \\
 & (this work)         & (Condon et al. 1998) \\
\midrule
 Tololo 1247-232      & 4.57$\pm$0.25  & 3.4$\pm$0.5          \\
 J125030-233323\textdagger       & 72.61$\pm$0.74 & 69.9$\pm$2.8 \\
 J125001-233210       & 5.64$\pm$0.42 mJy & 4.6$\pm$0.5          \\
\bottomrule
\end{tabular}
\caption{Continuum flux density measured for Tololo 1247-232 and two other sources in the field.
The results are in agreement with previously published values of \citet{Condon1998}.
\textdagger A 2-component Gauss fit was performed for J125030-233323.}
\label{tab:21cmflux}
\end{table}

%%%%%%%%%%%%%%%%%%%%%%%%%%%%%%%%%%%%%%%%%%%%%%%%%%%%%%%%%%%%%%%%%%%%%%%
%%%%%%%%%%%%%%%%%%%%%%%%%%%%%%%%%%%%%%%%%%%%%%%%%%%%%%%%%%%%%%%%%%%%%%%
%%%%%%%%%%%%%%%%%%%%%%%%%%%%%%%%%%%%%%%%%%%%%%%%%%%%%%%%%%%%%%%%%%%%%%%
\section{Results}

%%%%%%%%%%%%%%%%%%%%%%%%%%%%%%%%%%%%%%%%%%%%%%%
%%%%%%%%%%%%%%%%%%%%%%%%%%%%%%%%%%%%%%%%%%%%%%%
\subsection{Lyman Continuum Emission in the HST/COS spectrum} \label{sec:LyCescape}
Using the individual data reduction steps as described in section \ref{sec:calcosmod}, we could
gradually decrease the noise level, in particular in the LyC part of the spectrum below 912\AA\ 
rest-frame. However, using the \texttt{CALCOS} standard extraction size of 63 pixel
(corresponding to the 2.5\,arcsec PSA\ or 2.6\,kpc), no significant LyC leakage could be detected at first.

We also investigate the impact of Milky Way Ly$\updelta$ absorption on our LyC spectrum. To do so,
we first measure the Ly$\upalpha$ equivalent width ($\sim$14\AA). Using the square root part of the ``Curve of Growth'',
we derive a hydrogen column density of $\sim3.66 \times 10^{20}\ cm^{-2}$, which translates into a Ly$\updelta$
equivalent width of $\sim0.41\AA$. Thus, Ly$\delta$ Milky Way absorption has a negligible influence on our measurement.

\begin{figure*}
\centering
   \includegraphics[width=1\textwidth]{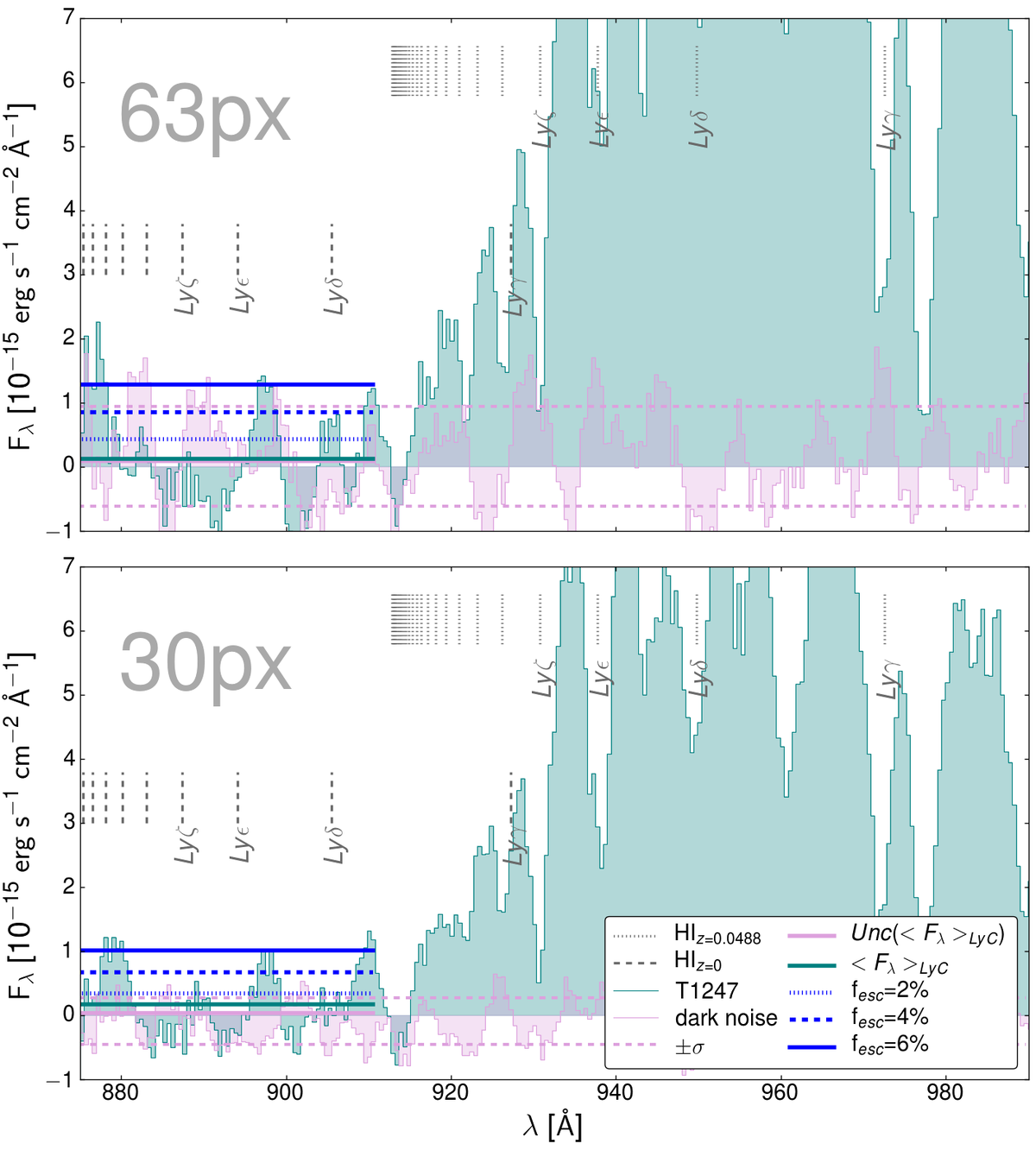}
     \caption{Spectral extractions of the rest-frame Lyman continuum part using the modified \texttt{CALCOS} pipeline.
     The spectra are extractions with apertures of 63 (top panel) and 30 pixel (bottom panel).
     Green bins show the signal of Tol1247 and red bins are extractions from a combined dark frame that was processed
     in exactly the same manner as the science frame, i.e. a superdark-corrected dark image (pure noise).
     That way, the typical dark noise level along the spectral extraction region
     could be estimated.
     The green horizontal line shows the mean flux level of the signal (Tol1247) in the range 875--911\,\AA. The red horizontal line
     is the same for the dark noise level (which should be close to zero). The blue horizontal lines indicate limits for
     the LyC escape fraction of 6\% (blue solid line), 4\% (blue dashed line) and 2\% (blue dotted line).
     The Lyman series and limits are shown for Tol1247 and the airglow as dotted and dashed vertical lines respectively.}
     \label{fig:LyC_bigap}
\end{figure*}

\begin{figure*}
\centering
   \includegraphics[width=1\textwidth]{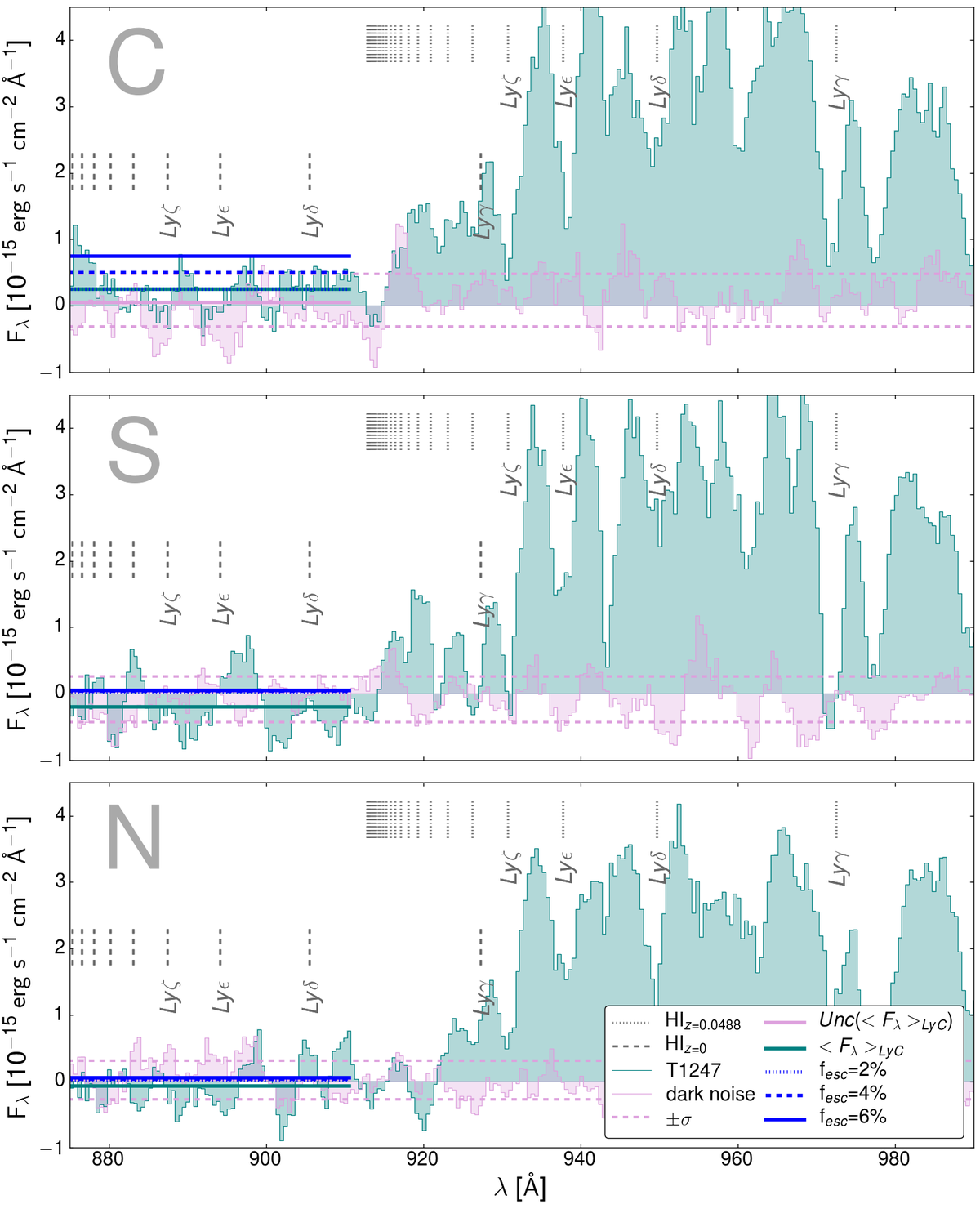}
     \caption{Same as Figure \ref{fig:LyC_smallap}, but for extraction heights of 16 pixel, centered on offset positions as
     shown in Figure \ref{fig:hstmaps}.}
     \label{fig:LyC_smallap}
\end{figure*}

Since LyC photons are mainly produced by massive clusters -- which are known from the ACS/SBC imaging to be extremely compact --
one expects LyC leakage mainly along the line-of-sight to these.
Hence, the LyC signal is most likely diluted as averaging over all extracted pixel rows is done.
For that reason, spectral extractions
with decreasing aperture sizes were performed, in expectation of an increase in signal-to-noise ratio (SNR) with a decreasing
extraction size. Figure \ref{fig:LyC_bigap} shows the resulting spectrum of the LyC part when
using the nominal extraction height of 63 px as well as an extraction height of 30 px around the center, corresponding to
$\sim$2.5arcsec\ (2.6\,kpc) and $\sim$1.2arcsec\ (1.3\,kpc) respectively. As shown in Figure \ref{fig:LyC_smallap},
we have also extracted spectra using a height of only 16px (which is 0.3\,arcsec or 0.3\,kpc), positioned northwards, central and southwards of the two main stellar clusters
(see regions N,C,S in the UVC image of Figure \ref{fig:hstmaps}).
It is found that smaller extraction regions centered around the stellar clusters indeed lead to a significant increase of
the SNR in the LyC, measured between 875 and 911\,\AA\ rest-frame. Table \ref{tab:LyC_extractions} summarizes all our LyC flux measurements.

Note that the noise levels (red curves in Figure \ref{fig:LyC_bigap} and \ref{fig:LyC_smallap}) are extracted from a combined dark frame that was processed
in exactly the same manner as the science frame, i.e. a superdark-corrected dark image (pure noise). That way, the typical dark noise level along the spectral extraction region
could be estimated.

The typical $\pm \upsigma$ noise levels shown in Figures \ref{fig:LyC_bigap} and \ref{fig:LyC_smallap} as red dashed lines,
are calculated from the standard deviation of the dark noise (rms$_{dark}$) when binned to 0.5\AA\ only. Since the observed  mean flux of the Lyman
continuum, <LyC>$_{obs}$, is measured over a wavelength range between 875 and 911\,\AA\ rest-frame, and thus covering more than 70 bins,
we calculate the resulting SNR, assuming the dark noise is purely random:
\begin{equation}\label{eqn:SNR}
SNR = \frac{<LyC>_{obs}}{\nicefrac{rms_{dark}}{\sqrt{bins}}}
\end{equation}

\begin{table*}
\centering
\begin{tabular}{c c c c c c c c}
\toprule
extraction  &    offset    &  region &          <LyC>$_{obs}$          &           Unc(<LyC>$_{obs}$)           &   SNR   &   <LyC>$_{predicted}$  &     $^\dagger$ f$_{esc,LyC}$ \\
height      &  from center &    name &                                 &                                &         &                        &                   [\%] \\
\midrule
16          &     13       &    N    &         -0.066                  &           0.036               & -2.6     &            0.9         &         <4.0	       \\
16          &    -19       &    S    &         -0.201                  &           0.039               & -3.0     &            0.9         &         <4.3	\\
16          &     -3       &    C    &          0.247                  &           0.045               &  3.6     &           12.5         &          2.0$\pm$0.4 \\
63          &      0       &         &          0.122                  &           0.089               & -0.5     &           21.3         &         <0.4 \\
30          &      0	   &         &          0.172                  &           0.041               &  6.2     &           17.0         &          1.0$\pm$0.2  \\
\bottomrule
\end{tabular}
\caption{Observed and predicted LyC flux densities in units of 10$^{-15}$ erg/s/cm$^2$/\AA\ for all spectra shown in
Figures \ref{fig:LyC_bigap} and \ref{fig:LyC_smallap}. The SNR is calculated as given by Equation \eqref{eqn:SNR}.
Using ionizing flux densities predicted through LaXs, the absolute LyC escape fractions f$_{esc}$ were calculated.
$^\dagger$ Our result contrast those of \citet{Leitherer2016}, who find a total LyC escape fraction of 4.5$\pm$1.2\%.
We find that the discrepancy is related to a \textit{negative flux issue} that
might occur at very low countrates when using \texttt{CALCOS 2.21d} (see also appendix \ref{sec:appendix}).}
\label{tab:LyC_extractions}
\end{table*}

Table \ref{tab:LyC_extractions} shows that the SNR indeed increases when smaller apertures are used, with the highest SNRs and LyC escape fractions
measured for central apertures of 30 and 16px. Since the SNR is calculated under the assumption of purely random noise (which is most likely not true),
the 1$\upsigma$ errors on the derived escape fractions are only lower limits on the real error. We conclude that the LyC escape fraction
of Tol1247 is 1.5$\pm$0.5\% only.

Our result contrast those of \citet{Leitherer2016}, who find a LyC escape fraction of 4.5$\pm$1.2\%.
We find that the discrepancy is related to a \textit{negative flux issue} that
might occur at very low countrates when using \texttt{CALCOS 2.21d} 
(see also appendix \ref{sec:appendix}).

%%%%%%%%%%%%%%%%%%%%%%%%%%%%%%%%%%%%%%%%%%%%%%%
%%%%%%%%%%%%%%%%%%%%%%%%%%%%%%%%%%%%%%%%%%%%%%%
\subsection{Constraints from UV Absorption Lines} \label{sec:uvabs}

Our COS spectra cover a multitude of UV absorption lines (see Figure \ref{fig:spectrum1327both}).
In order to study bulk motions
of the gas, we place all lines on a common velocity grid and create inverse variance weighted
average spectra tracing the neutral and ionized gas respectively.
We use the Si\,II transitions with ionization potentials below 1\,Rydberg at
$\uplambda\uplambda$ 1190,1193~\AA\ (CENWAVE=1327, segment B),
$\uplambda\uplambda$ 1304~\AA\ (CENWAVE=1327, segment A),
along with transitions of O\,I at $\uplambda\uplambda$ 1302~\AA\ and
C\,II at $\uplambda\uplambda$ 1334~\AA\
to trace kinematics of the neutral medium.
The Si\,IV lines
at $\uplambda\uplambda$ 1122~\AA\ (CENWAVE=1327, segment B)
and $\uplambda\uplambda$ 1393,1402~\AA\ (CENWAVE=1327, segment A)
were used as ionized gas tracer.

\begin{figure}
\centering
   \includegraphics[width=1\columnwidth]{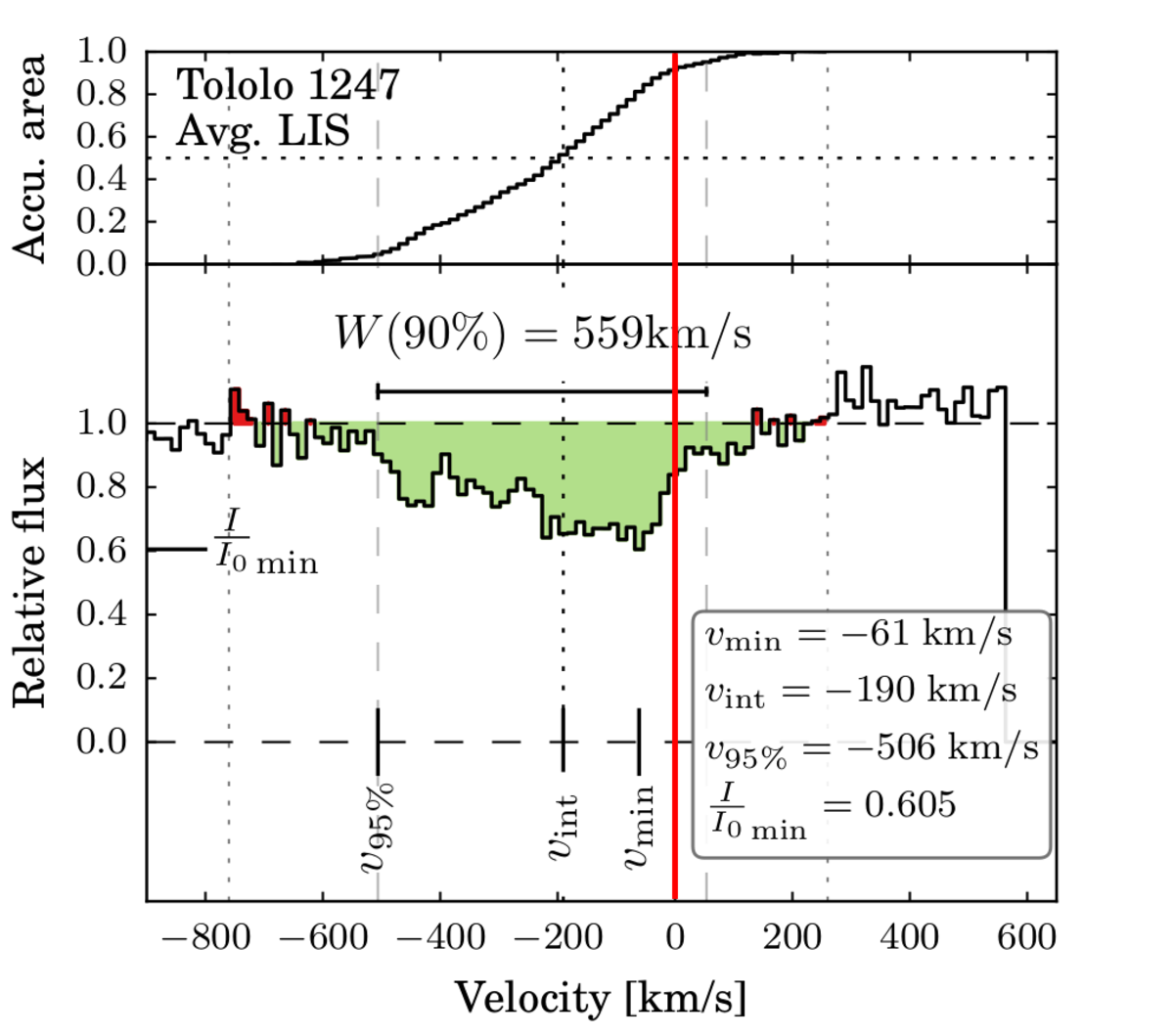}
     \caption{Mean low-ionization (LIS) line profile (containing Si II, O I and C II) with velocity parameters as defined by \citet{Rivera-Thorsen2015},
     where v$_{int}$ is the velocity which has 50\% of the integrated absorption on each side, v$_{95\%}$ is the
     absolute value of the velocity which has 95\% of the integrated absorption on its red side and 
     v$_{min}$ is the velocity at the minimum intensity, i.e. bulk of the gas moves with a velocity v$_{min}$.}
     \label{fig:dynvels}
\end{figure}

%\ref{fig:dynvels}
We calculate velocity parameters as defined by \citet{Rivera-Thorsen2015}, who
derived the same quantities for the \textit{Lyman Alpha Reference Sample}
(LARS; \citet{Hayes2013,Hayes2014,Ostlin2014}).
A summary of all derived quantities is shown in Table \ref{tab:LISquanti}.
As Figure \ref{fig:dynvels} shows, the linewidth $W(90\%)$ -- defined as the distance on the velocity
axis from 5\% to 95\% integrated absorption -- is very wide with a value of 559\,km/s and
the gas moves very fast, at a line-of-sight integrated
central velocity v$_{int}$ of -190\,km/s -- the velocity which has 50\% of the integrated absorption on
each side. Compared to the 14 LARS galaxies, Tol1247's values of W(90\%) and v$_{int}$ are found among the extremes,
slightly superseded only by two LARS galaxies.

Although most of the gas absorption is
blueshifted, also gas moving at positive velocities is seen and in particular a
sharp rise towards 0-velocity can be identified. This behavior is
very similar to the findings of \citet{Rivera-Thorsen2015}. However, with a
velocity v$_{95\%}$ of -506\,km/s -- the absolute value of the velocity which has 95\%
of the integrated absorption on its red side -- Tol1247 even outstands all of the LARS galaxies.
This indicates that very strong feedback processes are at work in Tol1247 that
accelerate the ISM. This facilitates the escape of Ly$\upalpha$ photons through superwinds, resembling the findings of
\citet{Duval2015}.
% Also makes gas unstable --> stronger accelleration --> RT instabilities
% Tol1247+LARS14 in Verhamme+ 2015 --> what were their conclusions?

\begin{figure}
\centering
   \includegraphics[width=1\columnwidth]{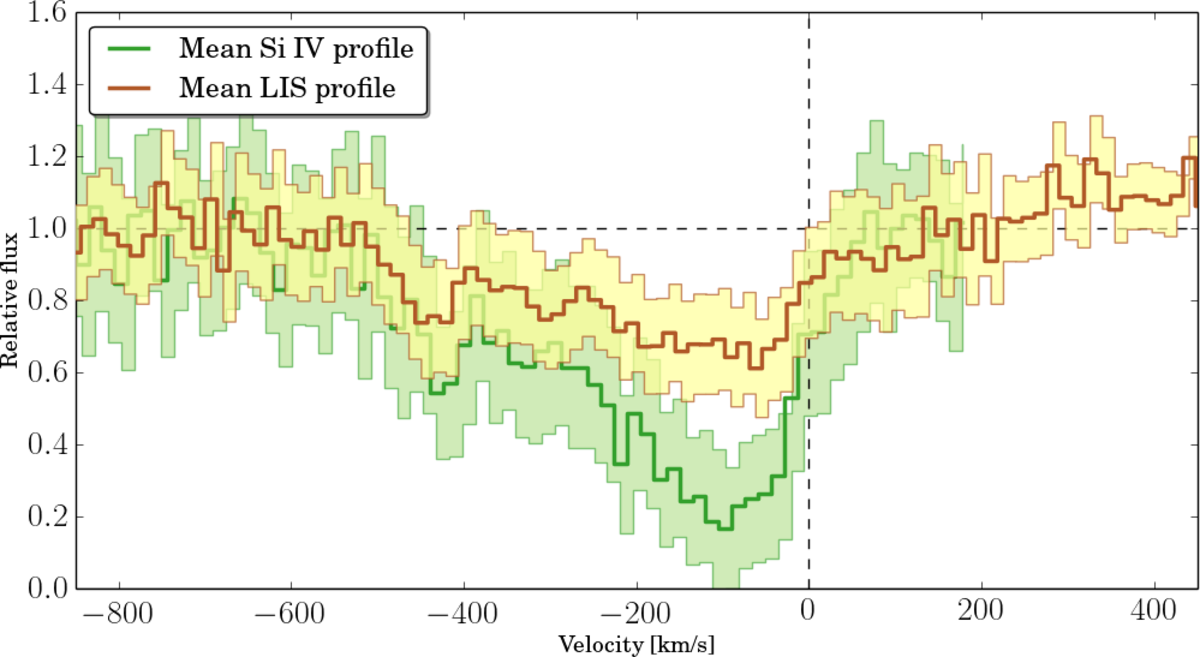}
     \caption{Comparison of the mean low-ionization (LIS) line profile (containing Si II, O I and C II) and the mean Si\ IV absorption profile.
      The overall trend is similar for both, the neutral and ionized medium traced by the
      LIS and Si IV profile respectively. However, bulk of the gas along the line-of-sight is highly ionized, supporting LyC leakage in case
      of density bound regions.}
     \label{fig:ionstage}
\end{figure}

%\label{fig:ionstage}
A comparison of the mean low-ionization (LIS) absorption profile and Si IV in Figure \ref{fig:ionstage} shows that the low and high ionization
lines are very similar in their structure, e.g. both share an absorption bump at $\sim-450$\,km/s.
However, Si IV absorption is stronger with a minimum relative flux $\nicefrac{I}{I_0}_{min}$ of only $\sim$0.2 at $\sim-100$\,km/s.
This suggests that most of the gas in the observed medium is highly ionized,
increasing the probability for LyC photons to escape, e.g. through density bounded regions.

\begin{table}
   \begin{tabular}{lll}
      \toprule
      parameter     &   value  & comment \\
                    &          &         \\
      \midrule
       $\nicefrac{I}{I_0}_0$      & 0.84$\pm$0.15    & relative flux at v=0\\
       $\nicefrac{I}{I_0}_{min}$  & 0.60$\pm$0.07    & minimum relative flux \\
       $W(90\%)$              & 559.8$\pm$0.1\,km/s    & linewidth from 5 to 95\% \\
                              &                  & integrated absorption \\
       $v_{95\%}$             & -506.2$\pm$0.1\,km/s   & velocity which has 95\% of \\
                              &                  & absorption on its red side \\
       $v_{int}$              & -190.4$\pm$0.1\,km/s   & velocity which has 50\% of \\
                              &                  & absorption on each side \\
       $v_{min}$              & -61$\pm$111\,km/s      & velocity at minimum \\
                              &                        & relative flux \\
      \bottomrule
   \end{tabular}
   \caption{Quantities derived from the average low ionization line profile as seen in Figure \ref{fig:dynvels}.}
   \label{tab:LISquanti}
\end{table}

%%%%%%%%%%%%%%%%%%%%%%%%%%%%%%%%%%%%%%%%%%%%%%%
%%%%%%%%%%%%%%%%%%%%%%%%%%%%%%%%%%%%%%%%%%%%%%%
\subsection{Covering Fraction and Column Density}
\begin{figure*}
\centering
   \includegraphics[width=1\textwidth]{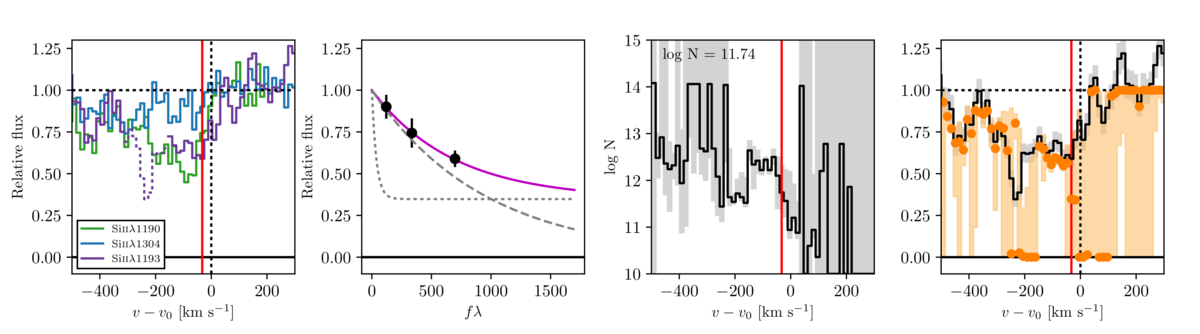}
     \caption{Explanation of the \textit{apparent optical depth method} as described in \citet{Rivera-Thorsen2015} and \citet{Rivera-Thorsen2017}.
     \textit{Left panel}: The Si~II lines utilized to solve for the column density and covering fraction (blended regions are drawn as dashed lines).
     For one velocity bin (marked with red line),
     the three measured relative intensities are shown in the \textit{second panel} as black markers in the f$\uplambda$ vs. I/I$_0$ plane.
     The best-fit function that finally leads to the column density and covering fraction is drawn in magenta. The solutions for N$_{Si II}$ and the
     covering fraction are shown in the \textit{third} and \textit{last panel} respectively.}
     \label{fig:AODmethod}
\end{figure*}

Next, we make use of the \textit{apparent optical depth method} as described by \citet{Savage1991}, used by
\citet{Pettini2002}, \citet{Quider2009} and more recently applied by \citet{Jones2013} and \citet{Rivera-Thorsen2015}, allowing to
compute the column density N$_{Si II}$ and covering fraction f$_c$ of these lines as a function of velocity offset from the
systemic zero-point (see Figure \ref{fig:AODmethod}).
The method utilizes the fact that the absorption depth of transitions from the same ground state, in the
case of an optical thin medium depend on the product of oscillator strength and wavelength $f \uplambda$, whereas an optical
thick medium with partial covering will yield identical absorption features in all transitions.
The ratio of the residual intensity I to the continuum I$_0$ within a given velocity interval
is utilized to calculate the covering fraction f$_c$ and the optical depth $\tau$:
\begin{equation}
    \frac{I}{I_0}\ =\ 1\ -\ f_c\ (1\ -\ e^{-\tau})
\end{equation}
The optical depth $\tau$ is further related to column density N$_{SiII}$ and the oscillator strength $f$ via
\begin{equation}\label{eqn:tauNrelation}
     \tau = f\uplambda\ \frac{\uppi e^2}{m_e c}\ N_{SiII}\ =\ f\uplambda\ \frac{N_{SiII}}{3.768\ \times\ 10^{14}},
\end{equation}
with the wavelength $\uplambda$ in \AA\ and N$_{Si II}$ in $cm^{-2}$.
Since we observed multiple Si lines that arise from the same energy level,
but have different values of $f \lambda$, a minimizing routine could be formulated to find
solutions for N$_{Si II}$ and $f_c$.
Practically, a grid of (N$_{Si II}$, $f_c$) pairs was defined and limited by physically sensible values only.
The squared residuals of subtracting the measured value of the ratio I/I$_0$ from the one found at the grid point,
could then be calculated and minimized.
%Confidence levels ($\chi^2 + 1 $) along the grid points for a single velocity bin
%are exemplary shown in Figure \ref{fig:confidence}.
We refer to \citet{Rivera-Thorsen2015} for a detailed description of the applied method.

Similar to the LARS galaxies, also Tol1247 shows no or little dependency of $f \uplambda$ on the absorption depth of the considered silicon lines.
Thus, the medium must be optically thick, but \textit{clumpy} since strong residual flux is seen at all velocities with a minimum
value (maximum covering) of $\nicefrac{I}{I_0}_{min}$ of $\sim$0.35.
The derived covering fraction as seen in right panel of Figure \ref{fig:AODmethod} shows that
even the maximum value for f$_c$ is low with a value of $\sim$0.7.
Note that f$_c$ basically follows the absorption feature.
For some of the spectral bins however, the derived errors of the
fitting procedure are spanning over the entire range of physical values,
indicating that no unique solution could be found.
This is due to the non-linear solution characteristics of the applied method.
%The confidence levels in the f$_c$ vs. N$_{Si II}$
%plane for e.g. the systemic velocity bin are given in Figure \ref{fig:confidence}, showing that f$_c$ could not be constrained, whereas for
%N$_{Si II}$ good estimates could still be derived.
From the observed low covering fraction, one might expect Ly$\upalpha$ emission at line center. However this is not observed (see Figure \ref{fig:lyaline}),
suggesting that an additional somewhat \textit{diffuse} component of H~I exists with a lower column density that may instead
absorb the Ly$\upalpha$ radiation.
Such low column density gas would still re-shape the Ly$\upalpha$ line without leaving an observable imprint in the silicon absorption lines.
Hence, we now investigate whether our observations are still consistent
with the presence of such a gas component, in particular at the systemic velocity, where we find that the
Si~II column density is lowest with an lower limit of only $\sim10^{11.3}\ cm^{-2}$ (see third panel of Figure \ref{fig:AODmethod}).
The metallicity of Tol1247 is 12+log(O/H)=8.1 \citep{Terlevich1993} or 1/4 Z$_{\odot}$. Thus, the Si~II column density can be converted
% Solar Reference Abundance for Si taken from:
% James Lequeux, The Interstellar Medium
% ISSN 0941-7834
% ISBN 3-540-21326-0 Springer Berlin Heidelberg New York
into a H~I column density of $\sim2.3\ \times\ 10^{16}\ cm^{-2}$ assuming 12+log(Si/H)$_{\odot}$=7.55 \
as given in \citet{Lequeux2005}
and that all Si in the neutral medium is in form of Si~II.
% however, this assumes that all Si is in form of Si II which is OK for normal galaxies, but probably not for TOL1247
% typical fractions can be found here: http://iopscience.iop.org/article/10.1086/308173/fulltext/
% The Astrophysical Journal, 528:310-324, 2000 January 1
This is indeed less than the typical value of $1.6\ \times\ 10^{17}\ cm^{-2}$ \citep{Mo2010} for
Lyman Limit Systems (LLS), supporting the escape of ionizing photons from the central region,
which is assumed to be co-spatial with the gas seen in absorption at systemic velocity.
At the same time, this is $\sim$10$^4$ optical depths for Ly$\upalpha$, hence absolutely optically thick to Ly$\upalpha$.
The presence of an additional \textit{diffuse} component thus might explain the lack of Ly$\upalpha$ flux at the systemic velocity,
in a system that at the same time leaks ionizing photons.
Farther out at a typical velocity $v_{min}$ at which bulk of the gas moves, the column density of Si~II is set to $\sim10^{12.5}\ cm^{-2}$, which translates
into a hydrogen column density of $3.6\ \times\ 10^{17}\ cm^{-2}$. Although still low, this is optically thick to both LyC and Ly$\upalpha$.
Thus, LyC likely escapes due to the combined effect of low gas column at systemic velocity (similar to a central cavity)
and the interstellar medium being clumpy, i.e. channels through which LyC can escape from the galaxy.

We further examine the overall N$_{Si II}$ sensitivity limit per velocity bin of our
observations, which can be identified through the error bars in Figure \ref{fig:AODmethod}. We see that the relative flux sensitivity is $\sim0.1$, so that for total covering
(diffuse medium): $\nicefrac{I}{I_0} = e^{-\uptau} = 0.9$. With the relation between the optical depth and column density (see Equation \ref{eqn:tauNrelation}),
and an average product of wavelength and oscillator strength of f$\uplambda$=384.8 for the given average spectrum,
a Si~II column density of N=$1.0\ \times\ 10^{11}\ cm^{-2}$
is found, describing the sensitivity limit of our observations. As above, this translates into a
hydrogen column density of $1.1\ \times\ 10^{16}\ cm^{-2}$, which is still optically thick to Ly$\upalpha$,
but optically thin by at least one magnitude to Lyman continuum; compare \citet{Verhamme2015}.
In agreement with the observations, we thus conclude that the observed medium is clumpy with optically thick clouds
that are partially covering. Additionally, a more diffuse component or inter-clump medium may be present
that is optically thick to Ly$\upalpha$, but provides at least one clear sightline for Lyman continuum to leak.
%
%\begin{figure}
%\centering
%   \includegraphics[width=1\columnwidth]{SiIILinesTololo1247.png}
%     \caption{The three Si II lines overlayed in velocity space for comparison. Pixels with contamination were masked out.}
%     \label{fig:Silines}
%\end{figure}

%\begin{figure}
%\centering
%   \includegraphics[width=1\columnwidth]{Confidence_N_vs_fc_mod.png}
%     \caption{The f$_c$ vs. N$_{Si II}$ solution plane at systemic velocity. Regions of high confidence are shown as green and yellow regions.
%     For the systemic velocity bin shown here, the applied method still allows to derive limits for the column density,
%     whereas the covering fraction could not be constrained.}
%     \label{fig:confidence}
%\end{figure}

%%%%%%%%%%%%%%%%%%%%%%%%%%%%%%%%%%%%%%%%%%%%%%%
%%%%%%%%%%%%%%%%%%%%%%%%%%%%%%%%%%%%%%%%%%%%%%%
\subsection{Lyman Alpha Emission Line Spectrum}
\begin{figure}
\centering
   \includegraphics[width=1\columnwidth]{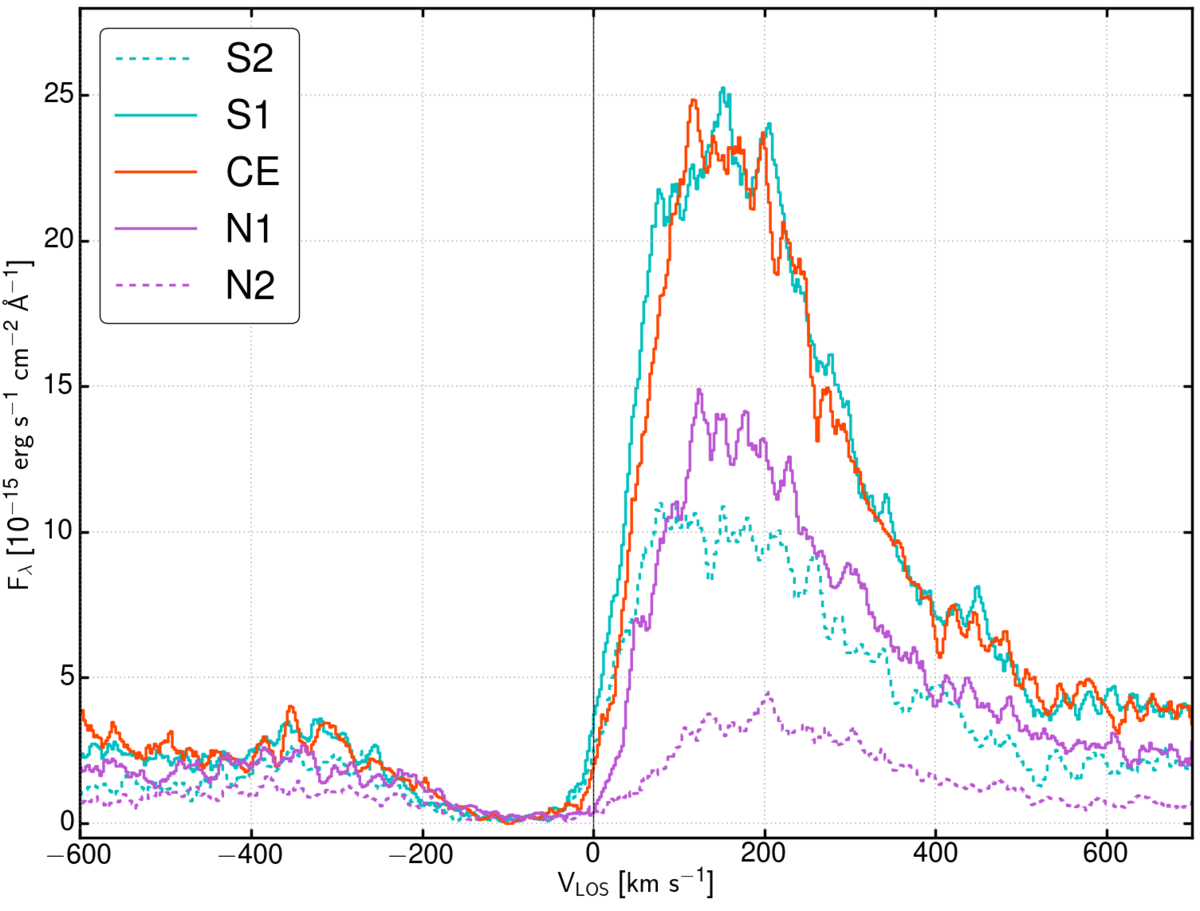}
     \caption{Lyman Alpha emission line profile extracted from North one quarter above (turquoise lines)
     to South one quarter (violet lines) below the center of the COS aperture.
     No obvious line shift can be deduced from the extractions. No flux is seen at systemic velocity,
     taking into the account the uncertainty of the measured redshift ($\pm$0.0002).}
     \label{fig:lyaline}
\end{figure}

%\ref{fig:lyaline}
The Ly$\upalpha$ profile in Figure \ref{fig:lyaline} shows extractions along the x-dispersion
around the central region of the COS aperture. Note that in this setup (CENWAVE=1327\AA) the nominal spectral extraction
height which corresponds to the 2.5\,arcsec PSA is 35 pixel only. We have performed four pixel wide extractions,
i.e. covering $\sim$0.29\,arcsec,
at positions of -8, -4, 0, +4 and +8 px offset from the PSA center (see also Figure \ref{fig:hstmaps}). Hence, we extract regions
ranging from 0.58\,arcsec North which is one quarter above (turquoise lines in Figure \ref{fig:lyaline}) to 0.58\,arcsec South
which is one quarter (violet lines) below the center of the COS aperture.
No obvious line shift can be deduced from the extractions, suggesting that both central clusters create
a common expanding gas shell.

The emission line shows a characteristic P-Cygni profile \citep{Kunth1998,Wofford2013,Rivera-Thorsen2015}, with
weak emission at the blue side.
Interestingly, the strongest metal absorption coincides with the Ly$\upalpha$ trough (compare Figures \ref{fig:lyaline} and \ref{fig:ionstage}),
probably meaning that the gas that absorbs most of the metals, also provides much of the absorption in Ly$\upalpha$.
%
%At first sight, the observed profile might be unexpected
%for a galaxy with a detected LyC leak as e.g. \citet{Verhamme2015} predict that good candidates for LyC leakage
%from a clumpy ISM should also have their Ly$\upalpha$ peak emission at the systemic redshift, which is not
%observed in Tol1247. However, as we have argued above, in the case of Tol1247 an additional diffuse component
%might absorb Ly$\upalpha$ at 0-velocity, but the medium is still optically thin for LyC photons.

%\begin{figure}
%\centering
%   \includegraphics[width=1\columnwidth]{LyACoverfracs.png}
%     \caption{Ly$\upalpha$ profile extracted from the whole PSA (\textit{top panel}) and covering fraction as in Figure \ref{AODmethod} (\textit{bottom panel}).
%     Interestingly, the strongest metal absorption coincides with the Ly$\upalpha$ trough,
%     probably meaning that the gas that absorbs most of the metals, also provides much of the absorption in Ly$\upalpha$.}
%     \label{fig:lya_covfrac}
%\end{figure}

%%%%%%%%%%%%%%%%%%%%%%%%%%%%%%%%%%%%%%%%%%%%%%%
%%%%%%%%%%%%%%%%%%%%%%%%%%%%%%%%%%%%%%%%%%%%%%%
\subsection{Radial Surface Brightness Profile}
At the given distance to Tol1247, the pixel scale of 0.04\,arcsec/px in Figure \ref{fig:hstmaps}
corresponds to 42 parsec per pixel. At this scale, complex structures can be revealed.
The H$\upalpha$ map shows an overall bar-like distribution with individual clumps of strong emission
-- likely associated with massive stellar clusters --
along its main axis and an increasingly filamentary structure towards the outer regions. In particular,
an arc with a diameter of roughly 300\,pc can be identified associated with the most luminous cluster
close to the center, indicative for a highly turbulent ionized medium or outflows.

\begin{figure}
\centering
   \includegraphics[width=1\columnwidth]{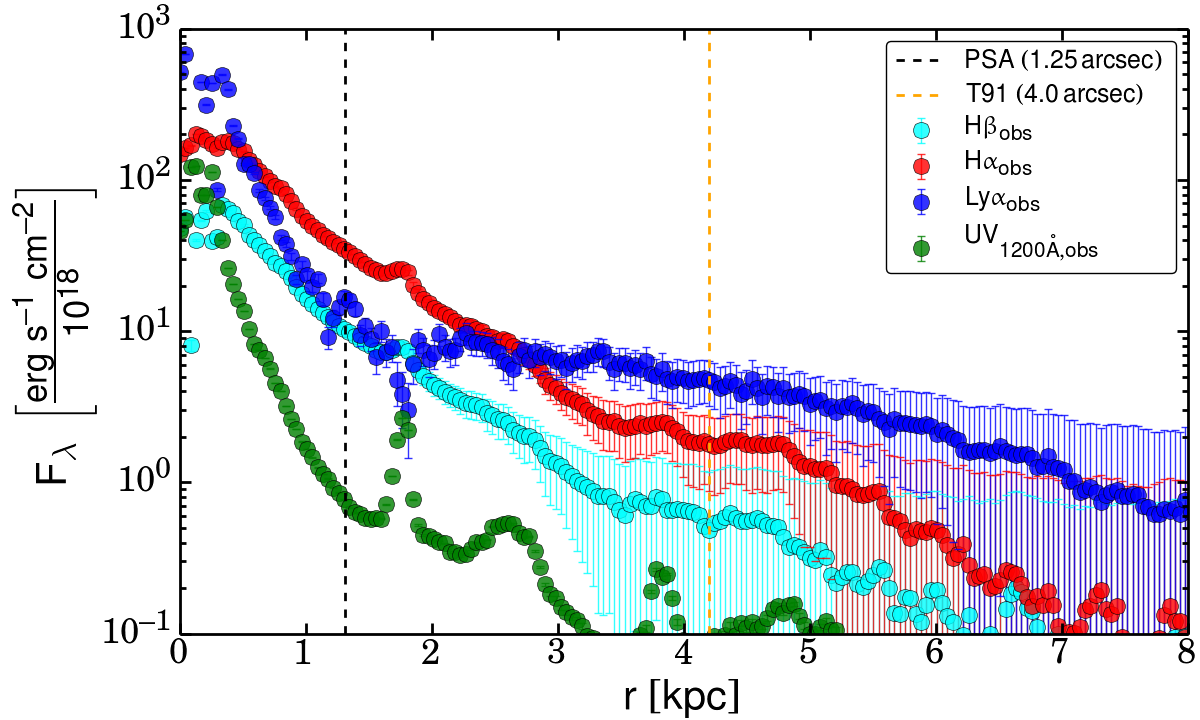}
     \caption{Mean surface brightness of H$\upalpha$, H$\upbeta$, Ly$\upalpha$ and UV continuum at $\sim$1200\AA\ evaluated from
     the center of the COS aperture outwards in steps of one pixel at a pixel scale of 0.04 arcsec/px, corresponding to 42 pc/px.
     The vertical black dashed line indicates the radius covered by the COS aperture and the vertical orange dashed line shows the
     limit of the (rectangular) aperture that was used by \citet{Terlevich1991}.}
     \label{fig:radprof_flux}
\end{figure}

The radial surface brightness profile in Figure \ref{fig:radprof_flux} shows that the
Ly$\upalpha$ emission is strongest in the central region, however with a steep decline up to a radius
of $\sim$1\,kpc. This is interesting, because such behavior may be explained by a low column density
or significant ISM clumping, i.e. less Ly$\upalpha$ scattering and a compact emission of Ly$\upalpha$.
Thus, we find another indication for LyC to escape directly.
% eventually add Verhamme+ in prep. or priv. comm.
Farther out, the Ly$\upalpha$ profile flattens and its surface brightness is less than the one in H$\upalpha$ up to
a radius of $\sim$3\,kpc.
At this point, the medium starts changing from mostly ionized to mostly neutral, until
at $\sim$5.2\,kpc no significant H$\upalpha$ emission is observed anymore. However, Ly$\upalpha$ is seen
up to $\sim$6.2\,kpc in form of a halo.
The latter one is likely produced due to the clumpyness of the ISM
in combination with strong ionizing flux emerging from the central clusters that is converted
to Ly$\upalpha$ that resonantly scatters outwards.
Due to the clumpyness the effect of the dust on the Ly$\upalpha$ line profile is less efficient than in homogeneous media,
causing an intense Ly$\upalpha$ emission that can escape even from a very dusty clumpy ISM \citep{Scarlata2009,Duval2015}.

%%%%%%%%%%%%%%%%%%%%%%%%%%%%%%%%%%%%%%%%%%%%%%%
%%%%%%%%%%%%%%%%%%%%%%%%%%%%%%%%%%%%%%%%%%%%%%%
\subsection{Extinction} \label{sec:extinction}
\begin{figure}
\centering
   \includegraphics[width=1\columnwidth]{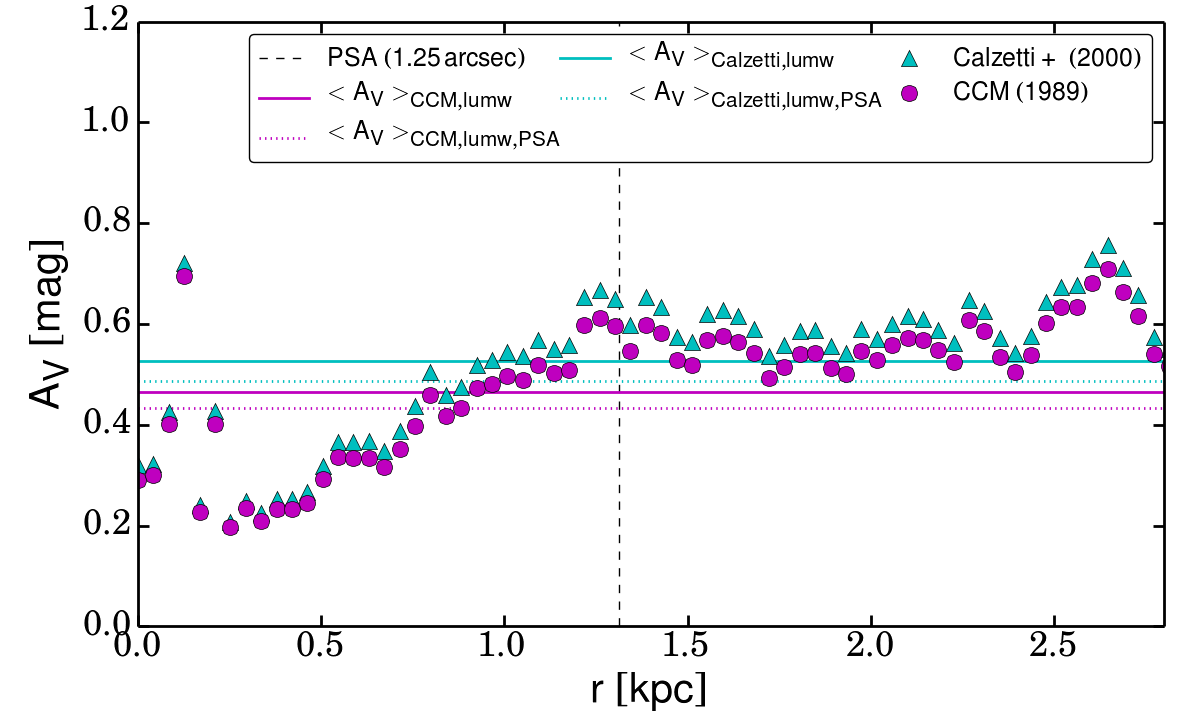}
     \caption{Average radial luminosity-weighted V-band attenuation within Tol1247 using the \citet{Cardelli1989} (\textit{magenta}) and \citet{Calzetti2000} law (\textit{cyan}).
     $A_V$ is evaluated from the center of the COS aperture outwards in annuli of one pixel at a pixel scale of 0.04 arcsec/px, corresponding to 42 pc/px.
     The vertical black dashed line indicates the radius covered by the COS aperture. Horizontal \textit{solid lines} are set at the average
     luminosity-weighted attenuation found within the whole galaxy (limited by H$\upbeta$ sensitivity) and the COS aperture (\textit{dashed lines}).}
     \label{fig:radprof_ext}
\end{figure}

% EXTINCTION
% Cardelli, Clayton, Mathis:
% E(B-V) for galaxy: 0.154683995013 mag
% E(B-V) in PSA: 0.143396439628 mag
%
% Calzetti:
% E(B-V) for galaxy: 0.130601964029 mag
% E(B-V) in PSA: 0.121071715589 mag
%
The Galactic dust reddening towards Tol1247 is $E(B-V)=0.075$\,mag or $A_V$=0.23\,mag \citep{Schlafly2011}.
Note that all observational fluxes reported are already corrected for the Galactic reddening and
all $A_V$ values only account for the intrinsic attenuation within Tol1247.
Attenuation due to dust within Tol1247 was calculated from the Balmer decrement using
the \citet{Cardelli1989} (CCM) attenuation law for the Milky Way (MW) and the \citet{Calzetti2000} one for starbursts (SB).
An intrinsic, theoretical $\nicefrac{H\upalpha}{H\upbeta}$ ratio of 2.80 was calculated for an electron temperature
of 12100\,K, which was published by \citet{Terlevich1993} based on the [O\,III]$\uplambda5007$/[O\,III]$\uplambda4363$
emission line ratio.
Both prescriptions give similar results for the average 
V-band attenuation of the galaxy, i.e. 0.47 and 0.53\,mag $A_V$ (see Figure \ref{fig:radprof_ext}) for MW and SB laws.
Within the COS aperture, the average attenuation $A_V$ is 0.43 and 0.49\,mag, respectively.\ 
%Low extinction agrees well with the measured stellar UV continuum slope in the PSA of $\upbeta_{PSA} = -2.1$
%that is very close to a dust-free value of young massive stars. 
Although the COS average is only slightly less than for
the whole galaxy, the attenuation in the innermost $\sim500\,pc$ is in fact much lower with $A_V$ down to $\sim$0.2.

Given the starburst nature of Tol1247, we use $A_V=0.49$ derived using the Calzetti law to estimate the average
total hydrogen ($N_H=N_{H I}+2N_{H_2}+N_{H II}$) column density $N_H$ within the COS aperture
using equation \eqref{eqn:coldens} taken from \citet{Guver2009}.
\begin{equation}\label{eqn:coldens}
N_H\ =\ 2.21\ \times\ 10^{21}\ \times\ A_V,
\end{equation}
where $N_H$ is given in $cm^{-2}$ and $A_V$ in mag.
Using this, a total hydrogen column density of $1.1\ \times\ 10^{21}\ cm^{-2}$ is found.
However, we are mainly interested in $N_{H\,I}$, since both LyC and Ly$\upalpha$ are absorbed or scattered
by the neutral atomic gas.
Thus, we consider an empirically motivated range of molecular-to-atomic hydrogen ratios ($\mu$)
as well as ionization fractions ($\iota$), in order to calculate lower and upper limits for $N_{H\,I}$:
\begin{equation}\label{eqn:coldensHI}
N_{H I}\ =\ \frac{N_H}{1 + \mu + \iota},
\end{equation}
where $\mu=\nicefrac{2 N_{H_2}}{N_{H I}}$ and $\iota=\nicefrac{N_{H II}}{N_{H I}}$.
For the upper limit atomic hydrogen column density we adopt $\mu=\nicefrac{1}{5}$, which is lower than the value that is typically found in
most present-day galaxies ($\nicefrac{1}{3}$). We have chosen this value based on the results of the COLD GASS survey \citep{Saintonge2011a,Saintonge2011b,Kauffmann2012},
which revealed that bluer galaxies tend to have lower molecular gas fractions\footnote{see: \url{http://www.iram-institute.org/EN/content-page-298-7-158-240-298-0.html}}.

Support for a lower molecular gas fraction in starbursts is also found by e.g. \citet{Amorin2016}, who studied gas fractions and depletion times
in a sample of $\sim20$ low-metallicity starbursts (blue compact dwarf galaxies), accounting for a metallicity-dependent
CO-to-H$_2$ conversion factor. However, huge uncertainties remain, and the contrary seems to be found in e.g. Haro 11 \citep{Pardy2016}.
We further assume that only half of the hydrogen atoms are ionized ($\iota=1$), although our deep Si\,IV absorption profile
gives rise to an even higher ionization fraction. That way, we find an upper limit $N_{H I} < 4.9\ \times\ 10^{20}\ cm^{-2}$.
To derive a lower atomic hydrogen column density we adopt $\mu=3$ and $\iota=3$, again most likely overshooting the true value and
we find a lower limit $N_{H I} > 1.6\ \times\ 10^{20}\ cm^{-2}$.
Even the lower limit is $\sim10^3$ optical depths for LyC photons, but as explained before
in the central region LyC photons could still escape through channels in the clumpy ISM.

%%%%%%%%%%%%%%%%%%%%%%%%%%%%%%%%%%%%%%%%%%%%%%%
%%%%%%%%%%%%%%%%%%%%%%%%%%%%%%%%%%%%%%%%%%%%%%%
\subsection{Total H$\upalpha$, H$\upbeta$ and Ly$\upalpha$ Flux Measurements}
Using our continuum-subtracted images and accounting for Galactic extinction towards Tol1247,
we calculate the total observed line fluxes for $H\upalpha$, H$\upbeta$ and Ly$\upalpha$.
The $H\upalpha$ and H$\upbeta$ fluxes were further corrected for internal extinction
using the \citet{Cardelli1989} and \citet{Calzetti2000} attenuation laws as described in the previous
section. Our results are summarized and compared to previous findings in Table \ref{tab:fluxes}.
Note that \citet{Terlevich1991} used an aperture of 8x8\,arcsec and thus should
recover bulk of the total flux of Tol1247.
Indeed, our $H\upalpha$ flux is comparable to the one published by \citet{Terlevich1991}, i.e.
only 4\% higher and thus within the uncertainty that was estimated by \citet{Terlevich1991}
to be $\sim$10\%. However, our total H$\upbeta$ flux is larger by $\sim$20\%.
Nevertheless, our extinction-corrected fluxes are still in good agreement with the ones published
by \citet{Rosa-Gonzalez2007} (that are based on \citet{Terlevich1991}).

\begin{table}
\centering
\begin{tabular}{l c c c l}
\toprule
Line & Flux$_{PSA}$               & Flux$_{tot}$               & $\frac{Flux_{tot}}{Flux_{ref}}$ & Ref. \\
(1)  & (2)                        & (3)                        & (4)                      &  (5) \\
\midrule
H$\upalpha$  & 25.9$\pm$0.3  & 52.6$\pm$4.1   & 1.04  & T91 \\
H$\upbeta$   &  8.0$\pm$0.2  & 16.2$\pm$1.2   & 1.20  & T91 \\
Ly$\upalpha$ & 28.0$\pm$0.5 &  64.1$\pm$6.4   & 1.00  & T93$\dagger$ \\
H$\upalpha_{0,MW}$  & 36.2$\pm$0.4  & 75.4$\pm$6.0  & 0.89 & RG07 \\
H$\upbeta_{0,MW}$   & 12.9$\pm$0.3  & 27.2$\pm$1.7  & 0.92 & RG07 \\
H$\upalpha_{0,SB}$  & 37.5$\pm$0.4  & 78.4$\pm$6.2   & 0.92 & RG07 \\
H$\upbeta_{0,SB}$   & 13.4$\pm$0.3  & 28.2$\pm$1.7   & 0.96 & RG07 \\
\bottomrule
\end{tabular}
\caption{Comparison of observed emission line fluxes and previously published values.
         The fluxes given in the table are corrected for Galactic extinction using E(B-V)=0.075.
         \textit{Column 1:} Line name.
         \textit{Column 2:} Continuum-subtracted line flux in units of 10$^{-14}$ erg/s/cm$^2$ within the PSA.
         \textit{Column 3:} Continuum-subtracted line flux in the same unit as before for the whole galaxy.
         \textit{Column 4:} Comparison to previously published values with the
         reference given in \textit{Column 5:}:
         T91: \citet{Terlevich1991}, T93: \citet{Terlevich1993}, RG07: \citet{Rosa-Gonzalez2007}.
         Fluxes corrected for extinction are indicated by an index $0$ followed by either MW
         or SB for the Milky Way or starburst attenuation law respectively.
         $\dagger$ The comparison here is based on a new calibration of the line flux in the IUE spectrum,
         that was first published by \citet{Terlevich1993}.}
\label{tab:fluxes}
\end{table}

%%%%%%%%%%%%%%%%%%%%%%%%%%%%%%%%%%%%%%%%%%%%%%%
%%%%%%%%%%%%%%%%%%%%%%%%%%%%%%%%%%%%%%%%%%%%%%%
\subsection{Star Formation Rates} \label{sec:SFR}
We calculate star formation rates (SFR) from 1.4GHz continuum as well as based on our MW and SB extinction-corrected
H$\upalpha$ and UV continuum fluxes. Stellar and nebular continuum are thereby related through:
$E(B-V)_{stellar}=0.44\ \times\ E(B-V)_{nebular}$ \citep{Calzetti2000}.
In order to make a fair comparison between all SFRs, we use SFR calibrations based on stellar population synthesis
performed by \citet{Sullivan2001}, which assume
a constant star formation history (SFH) and a \citet{Salpeter1955} initial mass function (IMF)
\footnote{Multiplication of our results by a factor 0.7 translates our SFRs to a \citet{Kroupa2001} IMF \citep{Bergvall2016}.}.
The conversions are shown in Equations \eqref{eqn:sfrha}--\eqref{eqn:sfr14100}.
Since UV and 1.4\,GHz calibrations undergo a strong evolution within the first $\sim$100\,Myr,
we calculate SFRs for ages of 10 and 100\,Myr. Our results are shown in Table \ref{tab:sfrcomp}.
\begin{equation}\label{eqn:sfrha}
      SFR_{H\upalpha}\ = \frac{L_{H\upalpha,0}\ [erg\ s^{-1}]}{1.22\ \times\ 10^{41}}\ [M_{\odot}\ yr^{-1}]
\end{equation}
\begin{equation}\label{eqn:sfruv10}
      SFR_{UV,10Myr}\ = \frac{L_{UV,0}\ [erg\ s^{-1}\ \AA^{-1}]}{3.75\ \times\ 10^{39}}\ [M_{\odot}\ yr^{-1}]
\end{equation}
\begin{equation}\label{eqn:sfruv100}
      SFR_{UV,100Myr}\ = \frac{L_{UV,0}\ [erg\ s^{-1}\ \AA^{-1}]}{5.76\ \times\ 10^{39}}\ [M_{\odot}\ yr^{-1}]
\end{equation}
\begin{equation}\label{eqn:sfr1410}
      SFR_{1.4GHz,10Myr}\ = \frac{L_{1.4GHz,NT}\ [erg\ s^{-1}\ Hz^{-1}]}{2.39\ \times\ 10^{27}}\ [M_{\odot}\ yr^{-1}]
\end{equation}
\begin{equation}\label{eqn:sfr14100}
      SFR_{1.4GHz,100Myr}\ = \frac{L_{1.4GHz,NT}\ [erg\ s^{-1}\ Hz^{-1}]}{8.85\ \times\ 10^{27}}\ [M_{\odot}\ yr^{-1}]
\end{equation}
The non-thermal luminosity $L_{1.4GHz,NT}$ at 1.4\,GHz rest must be given in $erg\ s^{-1}\ Hz^{-1}$, i.e.
we use our 1.4\,GHz flux measurement from Table \ref{tab:21cmflux}, convert it to cgs 
units and adopt a K-correction: $L_{1.4GHz,rest}\ =\ L_{1.4GHz,obs}\ \times\ (1\ +\ z)^{-\upalpha}$,
using a spectral power law slope $\upalpha = -0.8$. Next, the non-thermal fraction
at 1.4\,GHz is estimated using equations 21 and 23 of \citet{Condon1992}:
$L_{1.4GHz,rest,NT}\ =\ 0.87\ \times\ L_{1.4GHz,rest}$.     

%STAR FORMATION RATES 10Myr
%SFR from extinction corrected UVC using Cardelli law: 49.1805887156
%SFR from extinction corrected Halpha using Cardelli law: 34.7824542287
%
%SFR from extinction corrected UVC using Calzetti law: 46.9186999856
%SFR from extinction corrected Halpha using Calzetti law: 36.1619666874
%
%SFR from 1.4GHz continuum: 96.2695584142
%
%STAR FORMATION RATES 100Myr
%SFR from extinction corrected UVC using Cardelli law: 32.3556504708
%SFR from extinction corrected UVC using Calzetti law: 30.86756578
%
%SFR from 1.4GHz continuum: 25.9982197299
%
%STAR FORMATION RATE FROM 1.4GHz using
%empirical relation from Yun+ 2001
%and SFR_HA Calibration from Rosa-Gonzalez 2007
%SFR from extinction corrected Halpha using RG07 relation: 46.6780535749
%SFR from 1.4GHz using Yun+ 2001 empirical relation: 13.7708314593
%
\begin{table}
\centering
\begin{tabular}{l c c}
\toprule
                                              & 10Myr       & 100Myr \\
\midrule
SFR$_{H\upalpha,0,MW}$                        &  34.8       & 34.8 \\
SFR$_{H\upalpha,0,SB}$                        &  36.2       & 36.2 \\
SFR$_{UV,0,MW}$                               &  49.2       & 32.4 \\
SFR$_{UV,0,SB}$                               &  46.9       & 30.9 \\
SFR$_{1.4GHz}$                                &  96.2       & 26.0 \\
$\nicefrac{SFR_{H\upalpha,0,MW}}{SFR_{1.4GHz}}$   &  0.36       & 1.34 \\
$\nicefrac{SFR_{H\upalpha,0,MW}}{SFR_{UV,0,MW}}$  &  0.71       & 1.07 \\
$\nicefrac{SFR_{H\upalpha,0,SB}}{SFR_{1.4GHz}}$   &  0.38       & 1.39 \\ 
$\nicefrac{SFR_{H\upalpha,0,SB}}{SFR_{UV,0,SB}}$  &  0.77       & 1.17 \\
\bottomrule
\end{tabular}
\caption{Predicted SFRs in $M_{\odot}\ yr^{-1}$ for Tol1247, based on SF tracers and calibrations by \citet{Sullivan2001},
which assume a constant SFH and a Salpeter IMF. Values derived from extinction corrected fluxes are indicated by $0$.
Corrections based on a Milkyway and starburst attenuation law are denoted by $MW$ and $SB$ respectively.}
\label{tab:sfrcomp}
\end{table}

%%%%%%%%%%%%%%%%%%%%%%%%%%%%%%%%%%%%%%%%%%%%%%%
%%%%%%%%%%%%%%%%%%%%%%%%%%%%%%%%%%%%%%%%%%%%%%%
\subsection{Lyman Alpha Escape Fraction}
A typical H II region with a temperature of $\sim10^4$\,K and an electron density of 10$^3 cm^{-3}$,
is optically thick for energetic photons from massive O and B stars that are able to ionize
hydrogen (LyC photons: E>13.6\,eV or $\uplambda$<912\,\AA). However, 68\% \citep{Dijkstra2014} of these photons are
transformed into Ly$\upalpha$ radiation.
This is due to the recombination process that follows the absorption of a LyC photon.
In recombination theory, at least two limiting cases (A and B) can be distinguished for which
-- in the absence of dust -- the intrinsic line intensity ratio of Ly$\upalpha$ to H$\upalpha$ is known \citep{Osterbrock2006}.
In case A, the medium is optically thin in all hydrogen lines including the Lyman lines. Hence, Lyman
photons can freely escape, as well as the other photons from hydrogen recombination. Such
a H~II region is called \emph{density bounded}, because the luminosity is set by the density of the medium
rather than the number of photons. In case B, the medium is optically thick in all Lyman lines,
but optically thin in all other hydrogen lines, which is called \emph{ionization bounded}.
In this scenario, Ly$\upalpha$ photons are immediately absorbed (\textit{on the spot})
and thus excite another hydrogen atom that is followed by recombination and the emission
of Ly$\upalpha$. This process is called resonant scattering.
However, for both cases the line ratio of Ly$\upalpha$ to H$\upalpha$ ranges only from
around 7 to 12 for case B and A respectively \citep{Wofford2013}.
We now define a Ly$\upalpha$ escape fraction f$_{esc,Ly\upalpha}$
by dividing the measured line ratio to the intrinsic one. Following \citet{Hayes2015},
for the intrinsic Ly$\upalpha$ to H$\upalpha$ ratio a value of 8.7 is commonly adopted for case B recombination.
However, for Tol1247 we use an intrinsic ratio of 10, assuming that
the conditions in Tol1247 are at least partly density bounded, supported by our comparison
of the Si\,II and Si\,IV absorption profiles in Figure \ref{fig:ionstage} and
also suggested by the contour maps overplotted in Figure \ref{fig:hstmaps}
that follow the $\nicefrac{Ly\upalpha_{obs}}{H\upalpha_0}$ ratio, i.e. tracing the
Ly$\upalpha$ escape fraction. The contours show that the Ly$\upalpha$ escape fraction is highest
slightly offset from the line-of-sight to the stellar clusters. We argue that this is due to
the combined effect of a clumpy geometry and strong ionizing radiation field close to the central clusters,
which results in a density bounded emission line region.
Thus, we calculate the Ly$\upalpha$ escape fraction using
$f_{esc,Ly\upalpha}=\nicefrac{Ly\upalpha_{obs}}{(H\upalpha_{0}\ \times\ 10)}$ with H$\upalpha_0$ the
extinction corrected flux and find Ly$\upalpha$ escape fractions of
8.2$\pm$1.5\% and 7.5$\pm0.5$\% for the whole galaxy (including the halo) and within the COS aperture only.
Note that we find basically the same escape fraction for both attenuation laws (MW and SB).
%
%compute: 10.**(-0.4 * k1216 * EBV) == how much lya would escape if there were no such thing as scattering.
%may be able to claim that emission not heavily modulated by scattering?
%
%O32 ratio. 3.7. super high. as noted (Nakajima & Ouchi; Jaskot/Oey?) could be worth linking to Si IV / Si II absorption.

%%%%%%%%%%%%%%%%%%%%%%%%%%%%%%%%%%%%%%%%%%%%%%%
%%%%%%%%%%%%%%%%%%%%%%%%%%%%%%%%%%%%%%%%%%%%%%%
\subsection{Upper Limit for the atomic Gas Mass} \label{sec:atomicgas}
% source in VLA image right: NVSS J125001-233210; 1998AJ....115.1693C: 5.0 +/- 0.6 milliJy
% source in VLA image left: NVSS J125030-233323; 1998AJ....115.1693C: 71.0  +/- 2.8 milliJy
% Measurements in VLA data using CASA's imfit routine:
% NVSS J125001-233210: 5.64 +/- 0.42 mJy
% NVSS J125030-233323 (2 Gaussians were fitted): 19.52 +/- 0.37 mJy + 53.09 +/- 0.37 mJy = 72.61 +/- 0.74
% Tol1247-232 Continuum: 4.57 +/- 0.25 mJy
% 1 sigma noise level in collapsed H I line map: 0.28mJy/beam --> with 160km/s linewidth and 2sigma --> 90mJy km/s
The hydrogen hyperfinestructure line at 21\,cm remained undetected in Tol1247.
However with the current data, we can calculate an upper limit for the atomic gas mass. We do so by
measuring the noise level in the velocity-integrated (from -80 to +80km/s), continuum-subtracted
line map that is shown in the right panel of Figure \ref{fig:vlaimage} and find a 1$\sigma$
level of 0.28\,mJy/beam. Since Tol1247 is unresolved, the corresponding flux density is 0.28\,mJy. With a detection
threshold of 2$\sigma$ and a typical linewidth of 160\,km/s\footnote{
This estimate is based on the H I linewidth of two other local Ly$\upalpha$ emitters (LARS 1 and 5) with
comparable stellar masses, published by \citet{Pardy2014}.},
we estimate the 21\,cm velocity-integrated
flux limit S$_{21}$ to 90\,mJy\,km/s for Tol1247. With the widely used relation from \citet{Roberts1994},
which is shown in equation \eqref{eqn:mhi}, we arrive at an upper atomic mass limit of $\approx$10$^9 M_{\odot}$.
M$_{H I}$ is given in $M_{\odot}$, S$_{21}$ in mJy\,km\,s$^{-1}$ and D$_L$ in Mpc.
\begin{equation}\label{eqn:mhi}
      M_{H I} = 236 \times D_L^2 \times S_{21}\ [M_{\odot}]
\end{equation}

%%%%%%%%%%%%%%%%%%%%%%%%%%%%%%%%%%%%%%%%%%%%%%%%%%%%%%%%%%%%%%%%%%%%%%%
%%%%%%%%%%%%%%%%%%%%%%%%%%%%%%%%%%%%%%%%%%%%%%%%%%%%%%%%%%%%%%%%%%%%%%%
%%%%%%%%%%%%%%%%%%%%%%%%%%%%%%%%%%%%%%%%%%%%%%%%%%%%%%%%%%%%%%%%%%%%%%%
\section{Discussion}

%%%%%%%%%%%%%%%%%%%%%%%%%%%%%%%%%%%%%%%%%%%%%%%
%%%%%%%%%%%%%%%%%%%%%%%%%%%%%%%%%%%%%%%%%%%%%%%
\subsection{The H I Halo Size} \label{sec:halosize}
Due to resonant scattering, Ly$\upalpha$ emission is tracing the atomic gas in the galaxy.
From Figure \ref{fig:radprof_flux}, we find this to be $\sim$6.2\,kpc in radius.
Using our constraints on the atomic gas column density $N_{H I}$
(derived from the extinction A$_V$) and the upper limit atomic gas mass from the 21\,cm obersvations,
we now discuss whether the H I halo size seen through
Ly$\upalpha$ is tracing bulk of the atomic gas that is available to the galaxy or if additional H\,I
could be present that is not seen in the Ly$\upalpha$ radial profile.

In order to do so, a radial symmetric and homogeneous atomic gas distribution is assumed.
For a range of surface densities and halo sizes (=lengths over which the H I profile is constant),
atomic gas volume densities are calculated. The allowed parameter space is based on Figure 1 
of \citet{Bigiel2012}, who studied the neutral gas profiles in a set of 33 nearby spiral galaxies.
For the H I surface density we choose a range of 1--10\,M$_{\odot}\,pc^{-2}$ and we let the size
vary between 6--30\,kpc. For these conditions, the average H\,I volume density for Tol1247
can vary between 0.001 and 0.07\,cm$^{-3}$.

Next, we use the range of H\,I column densities $N_{H I}$ we have already derived in section
\ref{sec:extinction}. The maximum H\,I halo size is then found from the lowest volume
and column density,
at which the total atomic gas mass is just below the observed upper limit of $10^9\,M_{\odot}$.
That way, we find a maximum halo size of 13.6\,kpc with an average gas density of
0.0038\,cm$^{-3}$ and a H\,I column density $N_{H I}$ of $1.6\ \times\ 10^{20}\ cm^{-2}$.

This maximum value is 2.2 times the size of the Ly$\upalpha$ emitting area.
We stress that the maximum halo size was calculated based on the assumption that both, the
molecular and the ionized gas mass, are three times larger than the atomic mass.
Thus, Tol1247 contrasts five galaxies from the 
\textit{Lyman Alpha Reference Sample} (LARS; \citet{Hayes2013,Hayes2014,Ostlin2014})
that were observed in 21\,cm with the VLA \citep{Pardy2014}, because
in these LARS galaxies, the size of the 21\,cm emission is much larger than the size traced
through the Ly$\upalpha$ emission, i.e. by a factor of three or even more.

\citet{Pardy2014} further find an anti-correlation of the Ly$\upalpha$ escape fraction with H I mass.
With our upper atomic gas mass limit and the escape fraction as calculated before,
we find that also Tol1247 follows this trend.
Using the stellar mass $M_*=5 \times 10^9\,M_{\odot}$ published by \citet{Leitet2013}, we can further calculate an upper limit
for the gas fraction $f_{gas}=\nicefrac{M_{H I}}{M_*}<0.2$ and find that even our upper limit value
of $f_{gas}$ sets Tol1247 on the lowest end when comparing to LARS.
Only two LARS galaxies (LARS 8 and 9) out of 11 that
were detected in H I by \citet{Pardy2014} show values with $f_{gas}<0.2$; supporting the escape of
Lyman continuum and Ly$\upalpha$ photons in Tol1247 due to a lack of a huge H I halo and a low atomic gas fraction.

%%%%%%%%%%%%%%%%%%%%%%%%%%%%%%%%%%%%%%%%%%%%%%%
%%%%%%%%%%%%%%%%%%%%%%%%%%%%%%%%%%%%%%%%%%%%%%%
\subsection{Interpreting the SFRs derived from different Tracers}
Many different observational quantities from X-rays to the radio are related
to the star formation activity in galaxies \citep{Kennicutt1998a,Ranalli2003,Condon1992}.
However, the underlying physics
changes for each tracer, and so does also the timescale to which a tracer is
sensitive. Probably the most direct probe is UV emission
in the range between 1250 and 2000\,\AA\ that is created
by the youngest stellar population, i.e. O and B stars, with the latter ones
having a lifetime up to $\sim$300Myr.
A more recent tracer is H$\upalpha$, because only the most massive stars with lifetimes
of less than $\sim$20\,Myr produce sufficiently enough ionizing photons to
create H II regions from where H$\upalpha$ then emerges due to recombination.
This makes H$\upalpha$ almost independent of the star formation history (SFH).
Far infrared (FIR) emission is connected to star formation
due to heating of dust grains by UV photons, leading to thermal emission with
a peak in the wavelength range of 10 to 120$\upmu$m. The far infrared thereby
traces the average star formation on a timescale of $\sim$100Myr.
Another often used tracer is the non-thermal radio continuum.
Supernova remnants that occur after the first generation of most massive stars explode, i.e.
after $\sim$3.5\,Myr, accelerate relativistic electrons and produce synchrotron emission.
The spectral index of this non-thermal radio emission is -0.8, i.e. very steep compared
to the thermal spectral index of -0.1 which is attributed to H II regions.
In particular at lower frequencies, e.g. 1.4\,GHz, the thermal fraction for
normal star-forming galaxies is only 10\%,
making the 1.4\,GHz radio luminosity a good SFR tracer.
However, more than 90\% of this emission must be produced long after the individual
supernova remnants have faded out and the electrons have diffused throughout the galaxy \citep{Condon1992}.
For that reason, the radio synchrotron emission traces the star formation over a 
duration of $\sim$100\,Myr corresponding to the typical lifetime of relativistic
electrons in galaxies and making also the 1.4\,GHz luminosity-to-SFR-calibration strongly dependent on the SFH.

Common to all tracers is that their calibrations are very sensitive to the adopted IMF, the initial stellar mass
function \citep{Salpeter1955,Kroupa2001} and only SFRs calculated using the same assumptions, i.e.
SFH and IMF, can be compared to each other on a fair basis. In the case of a constant SFH in principle all tracers should give
equal results for ages of more than $\sim$100\,Myr; albeit complications due to e.g. dust attenuation introduce additional scatter.
Support for a constant SFH in most disk galaxies stems from the very tight relation
between FIR and radio luminosity \citep{Yun2001} that holds over many orders of magnitude.
Additionally, the existence of the main sequence (MS) of star-forming galaxies
\citep{Peng2010,Rodighiero2011,Sargent2012},
i.e. a tight positive relation between the SFR and stellar mass,
gives rise to continuously ongoing star formation.
Galaxies at the MS experience an equilibrium between gas accretion, star formation
and gas outflows \citep{Lilly2013}. Recently, the relation was also found to hold within individual galaxies
of the CALIFA survey \citep{Sanchez2012,Husemann2013} on a spatially resolved basis \citep{Cano-Diaz2016}.
And even galaxy populations observed at redshifts 2--3 when the Universe was only $\sim$2--3\,Gyr old,
follow the trend \citep{Genzel2010,Whitaker2012,Tacconi2013,Genzel2015,Schreiber2015}.
However, not all galaxies that form stars are located at the MS
(which can be parameterized following e.g. \citet{Whitaker2014}) and
starbursts are often classified as such by having a SFR that is four times more than the SFR on the
MS \citep{Rodighiero2011,Bergvall2016}. This is also true for Tol1247,
which has a SFR derived from H$\upalpha$ (see Table \ref{tab:sfrcomp}) that is $\sim$60 times the one on the main sequence
for the given stellar mass of $\sim5\ \times\ 10^{9}\ M_{\odot}$.

\citet{Rosa-Gonzalez2007} found that Tol1247 as well as other
H II galaxies in the \citet{Terlevich1991} sample are underluminous in the 1.4\,GHz continuum, or equivalent,
have a SFR derived from 1.4\,GHz that is lower than the SFR derived from H$\upalpha$.
Their SFR calibration was chosen such that in a well-defined control sample of starburst
galaxies the SFR derived from H$\upalpha$ agrees with the one derived through FIR (using the \citet{Kennicutt1998a}
calibration). The latter one was also used by \citet{Yun2001} when establishing the FIR-radio relation.
That way, based on extinction-corrected H$\upalpha$, the authors could make predictions
for the 1.4\,GHz luminosity at the \citet{Yun2001} relation.
The ratio between the observed and the predicted radio flux was defined by \citet{Rosa-Gonzalez2007} as \textit{d-parameter}.
Assuming that their conversion between H$\upalpha$ and FIR holds, the authors find that the galaxies
in the \citet{Terlevich1991} sample do not follow the \citet{Yun2001} relation, e.g. for Tol1247 the \textit{d-parameter}
was found to be 21\% only.
Given our extinction-corrected H$\upalpha$ and 1.4\,GHz measurements and using the SFR calibrations of \citet{Rosa-Gonzalez2007} we calculate
$SFR_{H\upalpha}=46.7$ and $SFR_{1.4GHz,NT}=13.8$, which confirms the results in \citet{Rosa-Gonzalez2007}.
The authors argue that the discrepancy in the sample with an average \textit{d-parameter} of $\sim$50\%
is due to the youth of the selected H II galaxies:
Most of the massive stars yet did not have enough time to evolve and
explode as supernovae. In this case, galaxies indeed are expected to have only very little synchrotron emission and
as a consequence their radio luminosities should be smaller than what one would expect
to find based on the SFR measured from other tracers such as H$\upalpha$.
With other words, in systems that are younger than $\sim$100\,Myr one has to be very careful when
adopting and comparing different SFR calibrations as they are tracing the star formation on different time scales,
leading to inconsistent results for the SFR.
We now ask whether the presence of a young population is in agreement with other data of Tol1247 that is available.
Beside the \textit{d-parameter}, further support comes from the large equivalent width EW(H$\upalpha$), which is
$\sim$530\AA\ \citep{Terlevich1991}. Assuming a constant star formation history, a value that high limits the age of the stellar population
to $\sim$30\,Myr (compare Figure 12 in \citet{Bergvall2016}).
\citet{Rosa-Gonzalez2007} also observed the radio continuum at 4.9 and 8.4\,GHz,
in order to derive the spectral index and distinguish between thermal and synchrotron radiation.
Interestingly, Tol1247's radio slope measured
between 4.9 and 8.4\,GHz is -0.85$\pm$0.24 and thus consistent with synchrotron emission rather
than thermal emission. Using our 1.4\,GHz flux we further re-measure the slope between 1.4 and 4.9\,GHz
and find -0.55$\pm$0.10. Hence, flattening of the slope is observed. We argue that the flattening
is due to free-free absorption that is proportional to $\sim \upnu^{-2}$ \citep{Condon1992} and thus
more efficient at lower frequencies. In any case unlike \citet{Rosa-Gonzalez2007} we do not find evidence
for a thermal radio slope in Tol1247. This is also found when using the relationship between H$\upalpha$
flux and thermal radio continuum flux density published by \citet{Caplan1986}.
After relating our extinction-corrected H$\upalpha$ flux to an expected thermal flux density at the
central observing frequency of 1.3522186 GHz and using $T_{e} = 12100\,K$, we find that the H$\upalpha$
flux implies a thermal flux density of 1.0\,mJy only, which is 22\% of the observed flux,
enforcing the non-thermally dominated nature of the continuum.
Thus, although Tol1247 seemingly follows the trend seen in bulk of the H II galaxies in the \citet{Terlevich1991} sample,
i.e. the 1.4\,GHz luminosity is deficient when compared to H$\upalpha$ due to the time-dependence of different SFR calibrations, the
spectral slope is consistent with synchrotron radiation rather than thermal emission.

\begin{figure}
\centering
   \includegraphics[width=1\columnwidth]{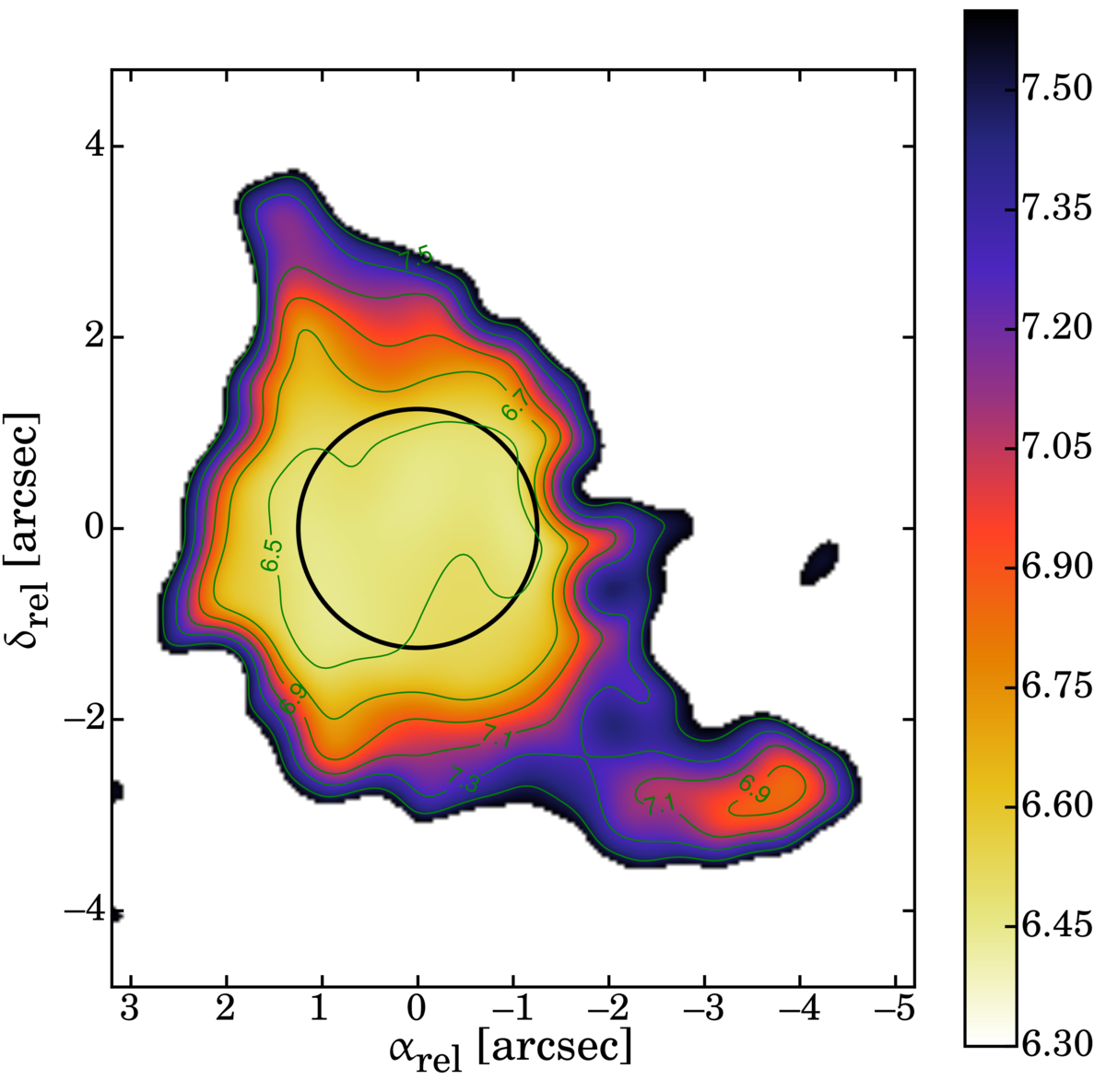}
     \caption{Derived stellar population ages of Tol1247. An age gradient can be identified with a dominating
     young population in the center and a radial increase in age. The maximum age of the galaxy is $\sim$30\,Myr.}
     \label{fig:age}
\end{figure}

We argue that the current data is in agreement with a star formation history lasting for $\sim$30\,Myr, which is likely more
complex than a continuously ongoing one and that
the observed low \textit{d-parameter} in Tol1247 is related to the time-depence of various SFR tracers.
This is supported by the H$\upalpha$ equivalent width and in broad agreement with the SFRs in Table \ref{tab:sfrcomp},
that we have derived using a set of calibrations from \citet{Sullivan2001} where all conversions (including the 1.4\,GHz)
are based on the same stellar population modeling, assuming a Salpeter IMF and constant SFH.
One can see that both, the UV and the 1.4\,GHz calibrations undergo a strong evolution within the first 100\,Myr and
that an age of $\sim$30\,Myr is reasonable, i.e. at this age likely all SFRs agree.
This result is further supported by our age map shown in Figure \ref{fig:age} that was produced as a result of our stellar population synthesis
using LaXs -- the Lyman alpha eXtraction software. The central region is dominated by
very young stars, i.e. a mean age of $\sim$3\,Myr is found within the COS aperture. However for the whole galaxy the
age is found to be $\sim$30\,Myr.

%%%%%%%%%%%%%%%%%%%%%%%%%%%%%%%%%%%%%%%%%%%%%%%%%%%%%%%%%%%%%%%%%%%%%%%
%%%%%%%%%%%%%%%%%%%%%%%%%%%%%%%%%%%%%%%%%%%%%%%%%%%%%%%%%%%%%%%%%%%%%%%
%%%%%%%%%%%%%%%%%%%%%%%%%%%%%%%%%%%%%%%%%%%%%%%%%%%%%%%%%%%%%%%%%%%%%%%
\section{Summary and Conclusion}
We have used the Cosmic Origins Spectrograph (COS) onboard the Hubble Space Telescope (HST)
to observe the Ly$\upalpha$ emission line, UV absorption lines (e.g. Si II) and the Lyman continuum
of the central region of Tol1247.
The data reduction procedure was optimized using a customized range of pulse height amplitude values,
filtering for orbital night data and an enhanced background subtraction routine. Furthermore, our modified
\texttt{CALCOS} version allows for spectral extractions of individual pixel rows along the x-dispersion.
Using these new features, in particular extracting only the central region that firmly
encloses the two stellar clusters, we could detect leaking LyC photons in a range of 875--911\AA.
We have further used the WFC3/UVIS and ACS/SBC onboard the HST to image the galaxy in seven filters
(broadband and narrowband).
This allowed us to apply LaXs -- the Lyman alpha eXtraction software \citep{Hayes2009} -- to 
perform stellar population synthesis and make predictions of the intrinsic Lyman continuum flux.
Using this, the LyC escape fraction was calculated to be\ 1.5$\pm$0.5\%, which
is significantly lower than the one recently found by \citet{Leitherer2016}.
We find that the discrepancy is related to the spectral extraction routine in \texttt{CALCOS 2.21d},
when a superdark frame is provided (see appendix \ref{sec:appendix}).
The Lyman continuum leakage is further supported by a multitude of other parameters:

\begin{itemize}
%%%%%%%%%%%%
% LIS and H IS absorption lines --> mostly ionized gas
\item{A comparison of the mean Si II and Si IV absorption profiles, suggests that most of the observed gas
is ionized, which facilitates LyC escape from density bounded regions.}

%%%%%%%%%%%%
% Hydrogen column density from Si II absorption lines
\item{With the Si II absorption lines considered, the observations are sensitive to a hydrogen column density of more than
$1.1\ \times\ 10^{16}\ cm^{-2}$, which is already optically thick to Ly$\upalpha$ (in agreement with the Ly$\upalpha$ emission profile
showing no emission at the systemic velocity),
but optically thin to Lyman continuum, giving rise to clear sightlines for Lyman continuum to leak.}

%%%%%%%%%%%%
% Common Lyman Alpha Shell
\item{Extractions of the Ly$\upalpha$ emission line along the x-dispersion do not show significant
differences in the shape of the P-Cygni profile, suggesting that both stellar clusters
that are located in the central region of Tol1247 share a common gas shell.}

%%%%%%%%%%%%
% Covering Fraction --> Clumpy
\item{From Si II absorption, we derive the covering fraction using the apparent optical depth method and
find that the gas clouds are optically thick. However, since residual flux is seen in all velocity bins, the ISM
must be sufficiently clumpy. Again a property supporting the direct escape of LyC photons.}

%%%%%%%%%%%%
% Halpha morphology
\item{Our H$\upalpha$ images reveal a rich
and complex morphology -- including arcs -- a sign for strong ongoing feedback processes accelerating
the ISM and presumably creating channels along the line of sight through which LyC photons escape.}

%%%%%%%%%%%%
% Extinction map
\item{The radial extinction profile reveals lines of sight of low extinction within the inner $\sim500\,pc$,
in agreement with a clumpy, anisotropic medium facilitating LyC photons to escape directly from
the innermost region of the galaxy.}

%%%%%%%%%%%%
% H I upper limit from VLA 21cm obs
\item{Tol1247 was imaged with the VLA in 21cm. The line was not detected, but a deep upper limit
atomic gas mass of $\lesssim10^9\,M_{\odot}$ was calculated. The gas fraction of Tol1247 is found
to be very low with $f_{gas} = \nicefrac{M_{H I}}{M_*}<0.2$.}

%%%%%%%%%%%%
% small H I halo
\item{A large fraction of the Ly$\upalpha$ emission in Tol1247 arises from a surrounding halo. We
find evidence that most of the atomic gas volume present in Tol1247 is traced by the observed Ly$\upalpha$ emission;
unlike what is seen in five galaxies from the LARS sample, where the halo seen in Ly$\upalpha$ makes
up only a small fraction of what is seen in VLA 21\,cm imaging.}

%%%%%%%%%%%%
% SFH
\item{Using a consistent set of SFR calibrations, we find that the SFRs derived from extinction-corrected H$\upalpha$, UV and
1.4\,GHz are in good agreement with a roughly continuously ongoing star formation during the last $\sim$30\,Myr. This
value also agrees well with the age derived from the stellar population synthesis.}

\end{itemize}

\section*{Acknowledgements}
% Referee
We thank the anonymous reviewer for his/her thorough comments and
suggestions.
% Pipeline
We are very thankful to C. Leitherer and S. Hernandez for providing us
with the code of the new \texttt{CALCOS} pipeline and their expertise.
% Matt Funding
M.H. acknowledges the support of the Swedish Research Council (Vetenskapsr{\aa}det) and
the Swedish National Space Board (SNSB), and is Fellow of the Knut and Alice Wallenberg Foundation.
% HST
This paper is based on observations made with the NASA/ESA Hubble Space Telescope, obtained at the Space Telescope Science Institute,
which is operated by the Association of Universities for Research in Astronomy, Inc., under NASA contract NAS 5-26555.
These observations are associated with program \#13027.
% VLA
We further made use of the Karl G. Jansky Very Large Array, a NRAO instrument.
The National Radio Astronomy Observatory is a facility of the National Science Foundation
operated under cooperative agreement by Associated Universities, Inc.
% NED
This research made use of
the NASA/IPAC Extragalactic Database (NED), which is operated
by the Jet Propulsion Laboratory, Caltech, under contract with NASA. 

%%%%%%%%%%%%%%%%%%%%%%%%%%%%%%%%%%%%%%%%%%%%%%%%%%



\bibliographystyle{mnras}
\bibliography{lycref} % if your bibtex file is called example.bib

%%%%%%%%%%%%%%%%%%%%%%%%%%%%%%%%%%%%%%%%%%%%%%%%%%

%%%%%%%%%%%%%%%%% APPENDICES %%%%%%%%%%%%%%%%%%%%%

\appendix

\section{Ad LyC escape fraction}\label{sec:appendix}

The LyC escape fraction reported in this paper is significantly lower than the one recently found by \citet{Leitherer2016},
hereafter L16. It is important to note that this is not due to the modeled flux density to which the observed flux is compared.
In fact, the theoretical flux density reported in our Table \ref{tab:LyC_extractions} is only roughly half of the model flux density
reported by L16 ($5.05\ \times\ 10^{-14} erg/s/cm^2/\AA$).

We find that the discrepancy is related to the spectral extraction routine in \texttt{CALCOS 2.21d},
when a superdark frame is provided.
The routine seems to fail at very low flux levels that may include negative count rates
(which likely happens after the superdark is subtracted from a low-signal science frame).
We have experimented with spectral extractions along different detector regions such as the nominal
background regions, where only noise is expected to be seen. However, even in this case,
residual \textit{positive} flux was reported by \texttt{CALCOS 2.21d} and no negative values were seen
in the resulting spectrum.
Since this is not an expected behavior, we examined the source code until we could isolate and fix the
problem, which we then reported to the \texttt{CALCOS} team.

It is important to note that this is only related to \texttt{CALCOS 2.21d}, which is (to our knowledge)
an unofficial release of the pipeline, that will be made available to the community \citep{Leitherer2016}.
We are very thankful to C. Leitherer and S. Hernandez who provided us with the code and hope that with this
finding we can contribute to the overall success of HST/COS.

%%%%%%%%%%%%%%%%%%%%%%%%%%%%%%%%%%%%%%%%%%%%%%%%%%

% Don't change these lines
\bsp	% typesetting comment
\label{lastpage}
\end{document}